\documentclass[10pt, twocolumn,article]{IEEEtran}
\usepackage{etex}
\usepackage{tikz}

\usetikzlibrary{shapes,arrows}

\usetikzlibrary{shapes,arrows,trees,calc}

\usetikzlibrary{arrows,shapes,positioning,shadows,trees}

\tikzset{
  basic/.style  = {draw, text width=2cm, drop shadow, font=\sffamily, rectangle,},
  root/.style   = {basic, rounded corners=2pt, thin, align=center,
                   fill=green!30},
  level 2/.style = {basic, rounded corners=6pt, thin,align=center, fill=green!60,
                   text width=12em},
  level 3/.style = {basic, thin, align=left, fill=pink!60, text width=8.5em},
  level 4/.style = {basic, thin, align=left, fill=pink!60, text width=8.5em}
}

\usepackage[T1]{fontenc}
\usepackage{tikz-qtree}
\usepackage{epigraph}
\usepackage{adjustbox}
\usepackage{setspace}
\usepackage{cite}
\usepackage[latin1]{inputenc}
\usepackage{epsf}\epsfverbosetrue
\usepackage{graphics,epsfig,color}
\usepackage{graphicx,epsfig}
\usepackage{epsfig}
\usepackage{multirow}
\usepackage{slashbox}
\usepackage{alltt}
\usepackage{subfigure}
\usepackage{color}
 \usepackage{float} 
\usepackage{url}
\usepackage{graphicx}
\usepackage{verbatim} 
\usepackage{footnote} 
\usepackage{latexsym} 
\usepackage{amsmath}
\usepackage{sidecap} 
\usepackage{color}
\usepackage{wrapfig}
\usepackage{algorithm}
\usepackage{eso-pic}
\usepackage{fix-cm}
\usepackage{mcite}

\usepackage{color}
\usepackage{wrapfig}
\usepackage{algorithm}
\usepackage{layout}
\usepackage{amsmath}
\usepackage{textcomp}
\usepackage{multirow}
\usepackage{soul,xcolor}
\usepackage{amsmath}
\usepackage{verbatim}
\usepackage{smartdiagram}
\usepackage{metalogo}
\usepackage{amsfonts}

\usepackage{pifont}
\usepackage{lscape} 

\usepackage{tikz}
\usetikzlibrary{shapes}
\usepackage{amsmath}
\usepackage{xspace}

\usepackage{tikz}
\makeatletter 
\@namedef{color@1}{red!40}
\@namedef{color@2}{green!40}   
\@namedef{color@3}{blue!40} 
\@namedef{color@4}{cyan!40}  
\@namedef{color@5}{magenta!40} 
\@namedef{color@6}{yellow!40}    

\newcommand{\graphitemize}[2]{%
\begin{tikzpicture}[every node/.style={align=center}]  
  \node[minimum size=5cm,circle,fill=gray!40,font=\Large,outer sep=1cm,inner sep=.5cm](ce){#1};  
\foreach \gritem [count=\xi] in {#2}
{\global\let\maxgritem\xi}  
\foreach \gritem [count=\xi] in {#2}
{%
\pgfmathtruncatemacro{\angle}{360/\maxgritem*\xi}
\edef\col{\@nameuse{color@\xi}}
\node[circle,
     ultra thick,
     draw=white,
     fill opacity=.5,
     fill=\col,        
     minimum size=3cm] at (ce.\angle) {\gritem };}%
\end{tikzpicture}  
}%

\newcommand{\cmark}{\ding{51}}%
\newcommand{\xmark}{\ding{55}}%
\newcommand{\mub}[1]{\textcolor{black}{#1}}

\newcommand{\drmub}[1]{\textcolor{black}{#1}}

\newcommand{\mhr}[1]{\textcolor{black}{#1}}

\newcommand*{\rom}[1]{\expandafter\@slowromancap\romannumeral #1@}
\newcommand{\Rmnum}[1]{\expandafter\@slowromancap\romannumeral #1@}
\makeatother

\begin{document}

\title{Software Defined Networks based Smart Grid Communication: A Comprehensive Survey}

\author{Mubashir~Husain~Rehmani, Alan Davy, Brendan Jennings, and Chadi Assi
\thanks{This publication has emanated from research supported in part by a research grant from Science Foundation Ireland (SFI) and is co-funded under the European Regional Development Fund under Grant Number 13/RC/2077. Part of this work has been done when M.H. Rehmani was doing Post Doctorate at Telecommunications Software and Systems Group, Waterford Institute of Technology, X91 K0EK Waterford, Ireland. \textit{(Corresponding author: M.H. Rehmani.)}}
\thanks{M.H. Rehmani is with Department of Computer Science, Cork Institute of Technology, Rossa Avenue, Bishopstown, Cork, Ireland. (e-mail: mshrehmani@gmail.com).}
\thanks{A. Davy is with the Telecommunications Software and Systems Group, Waterford Institute of Technology, X91 K0EK Waterford, Ireland (e-mail: adavy@tssg.org).}
\thanks{B. Jennings is with the Telecommunications Software and Systems Group, Waterford Institute of Technology, X91 K0EK Waterford, Ireland (e-mail: bjennings@ieee.org).}
\thanks{C. Assi is with the CIISE Department, Concordia University, Montreal, QCH3G 1M8, Canada  (e-mail: assi@ciise.concordia.ca).}
}

\maketitle

\begin{abstract}
\drmub{The current power grid is no longer a feasible solution due to ever-increasing user demand of electricity, old infrastructure, and reliability issues and thus require transformation to a better grid a.k.a., smart grid (SG). The key features that distinguish SG from the conventional electrical power grid are its capability to perform two-way communication, demand side management, and real time pricing. Despite all these advantages that SG will bring, there are certain issues which are specific to SG communication system. For instance, network management of current SG systems is complex, time consuming, and done manually. Moreover, SG communication (SGC) system is built on different vendor specific devices and protocols. Therefore, the current SG systems are not protocol independent, thus leading to interoperability issue. Software defined network (SDN) has been proposed to monitor and manage the communication networks globally. By separating the control plane from the data plane, SDN helps the network operators to manage the network flexibly. Since SG heavily relies on communication networks, therefore, SDN has also paved its way into the SG. By applying SDN in SG systems, efficiency and resiliency can potentially be improved. SDN, with its programmability, protocol independence, and granularity features, can help the SG to integrate different SG standards and protocols, to cope with diverse communication systems, and to help SG to perform traffic flow orchestration and to meet specific SG quality of service requirements. This article serves as a comprehensive survey on \drmub{SDN-based SGC}. In this article, we first discuss taxonomy of advantages of \drmub{SDN-based SGC}. We then discuss \drmub{SDN-based SGC} architectures, along with case studies. Our article provides an in-depth discussion on routing schemes for \drmub{SDN-based SGC}. We also provide detailed survey of security and privacy schemes applied to \drmub{SDN-based SGC}. We furthermore present challenges, open issues, and future research directions related to \drmub{SDN-based SGC}. }
\end{abstract}

\begin{IEEEkeywords}
Software defined network (SDN), smart grid (SG), advanced metering infrastructure (AMI), renewable energy resources (RERs), home area networks (HANs), network management.
\end{IEEEkeywords}

\section{Introduction}    
\label{sec:in}

\subsection{Motivation: Need of Software Defined Networks based SG}
\label{sec:motiv}

The conventional electrical power grid is undergoing a massive change. In this conventional power grid, electricity is typically generated through fossil-fuel based power generation units (e.g., nuclear, hydro, and coal based power generation units) and then transmitted it to the consumers via a huge network of transmission lines \cite{FaMX14}. Moreover, the flow of electric power is unidirectional i.e., from generation units to the consumers. With the ever-increasing user demand of electricity, old infrastructure, reliability issues, and prominence of renewable energy resources (RERs), the conventional power grid is \mub{no longer a} viable solution and thus require \mub{transformation to a better grid a.k.a., smart grid (SG).}   

\begin{figure*}[t] \centering
\includegraphics[width=0.9\textwidth]{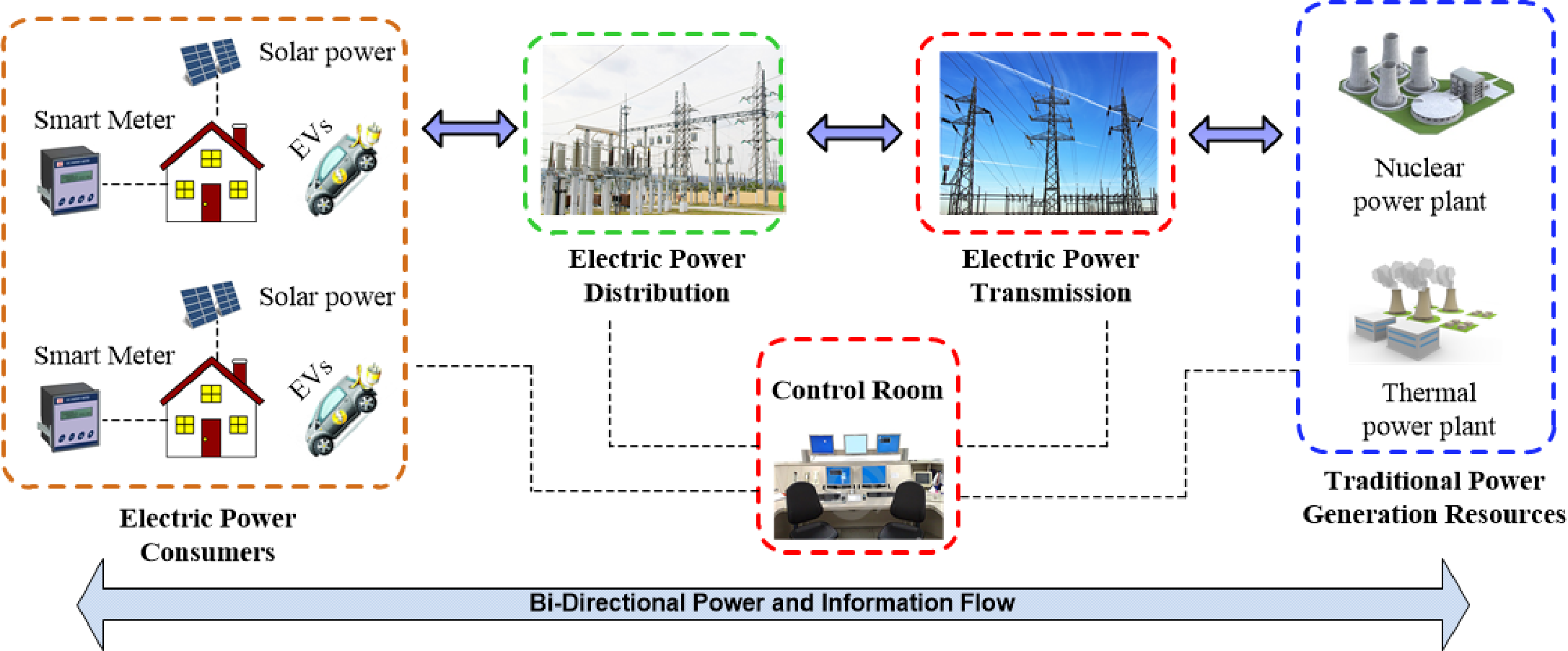}
\caption{\mub{Illustration of a typical \drmub{SG} architecture showing SG with its components such as consumers, PEVs, RERs and transmission and distribution network along with a control room which manages the bi-directional information and power flow within the SG.}} 
\label{fig:sgbasic}
\end{figure*}

The future \drmub{SG} will be equipped with advanced capabilities of automation, monitoring, and communication \cite{Rehmani17tii,Rehmani15access}. The key features that distinguish \drmub{SG} from the conventional electrical power grid are its capability to perform two-way communication, demand side management, and real time pricing. On top of that, SG will be insusceptible to faults and failures by using its self-healing capability. In addition, advanced metering infrastructure (AMI), plug-in electric vehicles (PEVs), supervisory control and data acquisition (SCADA), and \drmub{RERs} will be the indispensable parts of the SG. \mub{A typical \drmub{SG} architecture is illustrated in Fig.~\ref{fig:sgbasic}.}

Achieving a 100\% renewable future \drmub{SG} is one of the key focuses in many countries these days \cite{Kroposki17pem}. In this context, some roadmaps have been proposed for the integration of RERs into the \drmub{SG} \cite{Farhangi14pem}. It is envisaged that the future \drmub{SG} will completely rely on RERs by 2050 \cite{Mark17joule}. \drmub{RERs} are geographically dispersed and distributed in nature. They are disparate as they are gathered from different technologies. Moreover, RERs have less generation capacity compared to their counterpart i.e., traditional energy resources \cite{Yu11network}. Therefore, strong coupling of RERs into the \drmub{SG} requires timely and reliable communication \cite{Yu11network}. 

Interestingly, homes are the places where RERs are deployed at a massive level \cite{Heile10wcm}. In-house RERs, such as solar panels and even small scale distributed wind farms, which are connected with smart homes, can inject back substantial amounts of energy into the grid \cite{Zipperer13pieee}. This injection of energy by the smart homes into the \drmub{SG} needs to be supervised and conducted in a controlled manner for the overall stability of the power system. However, this supervision and monitoring of injected energy can only be done with the help of an efficient and reliable communication system. 

In smart homes \cite{Zipperer13pieee}, \drmub{consumers deal with a wide range of functions,} such as demand side management, real time pricing and billing, load scheduling, as well as surplus generated energy to the electricity suppliers. Additionally, software upgrades of smart meters are frequently required, through gateways to a number of smart meters, without the need to visit every meter location. Information exchanges can occur in the form of meter readings taken from meters to the utility, from meters to the AMI, and from the AMI to the utility. These information exchanges are either on-demand, scheduled a priori, or in the form of bulk transfers. Real-time pricing and time of use (TOU) pricing information exchanges also occur between the utility and the
meters. Again, these information exchanges need to be conducted through a secure and reliable communication system and need to be managed globally. 

\drmub{Though traditional approaches, like \drmub{multiprotocol label switching (MPLS)}, has been adopted initially by the utilities for \drmub{SG} communication system but it is not completely sufficient. One primary reason for not adopting MPLS in SG is because with the addition of new services in SG, the MPLS based routers need to be re-configured each time, resulting in disruption of services provided by the utilities \cite{Sydney13tsg}. Thus, one alternative that comes in our mind is software defined network (SDN), which further led to the emergence of \drmub{SDN} based \drmub{SG communication} \drmub{(SDN-based SGC)}.}

\drmub{SDN has been proposed to monitor and manage the communication networks globally. SDN revolutionized the way the communication network managed previously. The applicability of SDN in different domains is not new. SDN has been applied to data centers \cite{Singh17jpdc,Cui13jsac}, wide area networks (WAN) \cite{Ahmed14commag}, enterprises \cite{Chen16tetc,Lorenz17commag}, optical networks \cite{Thya16comst}, wireless networks \cite{Haque16comst}, wireless sensor networks \cite{Kobo17access}, and under water sensor networks (UWSN) \cite{Akyildiz16adhocnet}. Therefore, SDN paradigm with its capability to \drmub{separate} the control plane from the data plane can be broadly used as a basis for SG communication support.}

More precisely, since the SG relies heavily on communication networks for control, SDN can be employed to manage the communication entities in the SG system. By applying SDN in SG systems, efficiency and resiliency can potentially be improved~\cite{Dorsch17ceiesi}. For instance, the SDN based SG can be used for load balancing and shifting, for dynamically adjusting the routing paths for SG control commands \cite{Zhao2016}, fast failure detection \cite{Dorsch16globecom}, security \cite{Ghosh17icdcsw}, self-healing \cite{Lin18tsg}, and for monitoring and scheduling of critical SG traffic flows. Moreover, SDN will help to evolve the SG to embrace with new technologies, services, and needs.

\subsection{Contribution of This Survey Article}
In this survey article, we make the following contributions:
\begin{itemize}
\item We survey and classify advantages of SDN-based \drmub{SGC.}
\item We discuss SDN-based \drmub{SGC} architectures, along with case studies.
\item We provide an in-depth discussion on routing schemes for SDN-based \drmub{SGC}.
\item We provide detailed survey of security and privacy related schemes applied to SDN-based \drmub{SGC}.
\item We outline open issues, challenges, and future research directions related to SDN-based \drmub{SGC}.
\end{itemize}

\begin{table}[t]\footnotesize
\centering
\caption{List of acronyms and corresponding definitions.}
\label{acronym}
\label{tb1}
\begin{tabular}{|p{1.5cm}|p{5.5cm}|}
\hline
\bfseries Acronyms & \bfseries Definitions \\
\hline
AMI & Advanced Metering Infrastructure \\
\hline
CRSG & Cognitive Radio Smart Grid \\
\hline
DR & Demand Response \\
\hline
EBI & East Bound Interface \\
\hline
EI & Energy Internet \\
\hline
G2V & Grid-to-Vehicle \\
\hline
HAN & Home Area Network \\
\hline
HPEVs & Hybrid Plug-in Electric Vechiles \\
\hline
ICT & Information and Communication Technologies \\
\hline
IDS & Intrusion Detection System \\
\hline
IEC & International Electrotechnical Commission \\
\hline
IoT & Internet of Things \\
\hline
M2M & Machine to Machine \\
\hline
MPLS & Multiprotocol Label Switching \\
\hline
NAN & Neighborhood Area Network \\
\hline
NBI & North Bound Interface\\
\hline
NFV & Network Function Virtualization \\
\hline
PDC & Phasor Data Concentrator \\
\hline
PEVs & Plug-in Electric Vehicles \\
\hline
PMU & Phasor Measurement Unit \\
\hline
RERs & Renewable Energy Resources \\
\hline
REST & Representational State Transfer \\
\hline
SBI & South Bound Interface \\
\hline
SCADA & Supervisory Control And Data Acquisition \\
\hline
SG & Smart Grid \\
\hline
SDN & Software Defined Network \\
\hline
UWSN & Under Water Sensor Network \\
\hline
V2G & Vehicle-to-Grid \\
\hline
VPN & Virtual Power Plant \\
\hline
WAN & Wide Area Network \\
\hline
WBI & West Bound Interface \\
\hline
\end{tabular}
\end{table}

\subsection{Article Structure}
A list of acronyms used throughout the paper is presented in
Table~\ref{acronym}. The rest of the paper is organized as follows: \mub{Comparison with related survey articles is presented is Section~\ref{comparsur}.} In Section \ref{sgbk}, we discuss basics of SDN and SG. In the same section, we \mub{also highlight motivation} of employing SDN in SG. Moreover, case studies on the use of SDN in SG is also presented in this section. In Section \ref{adv}, we discuss the taxonomy of advantages of \drmub{SDN-based SGC}. Architectures for \drmub{SDN-based SGC} are discussed in Section~\ref{arc}. In Section \ref{rout}, routing schemes for \drmub{SDN-based SGC} are surveyed. Security and privacy schemes for \drmub{SDN-based SGC} are reviewed in Section \ref{secur}. Issues, challenges, and future research directions are mentioned in Section \ref{sec:openissues}. Finally, Section \ref{sec:conclusion} concludes the paper.

\begin{table*}[t]\footnotesize
\centering
\caption{Comparison of advanced metering infrastructure and field area networks. Neighborhood area networks are composed of AMI and FANs.}
\label{compamifans}
\begin{tabular}{|p{2cm}|p{6cm}|p{6cm}|}
\hline
\multicolumn{3}{|c|}{\bfseries Neighborhood Area Networks (NANs)} \\ \hline
       &  \bfseries  Advanced Metering Infrastructure (AMI)   &  \bfseries Field Area Networks (FANs)    \\ \hline
\bfseries Devices       &    Smart meters, which are used to monitor gas, electricity and water consumption   &  Devices which are used for fault detection, power grid protection, and/or control of the distribution grid     \\ \hline
\bfseries Communication       &   Two-way communications is required from/to consumers and utility    &   Two-way communications is required from/to consumers and utility    \\ \hline
\bfseries End Points       &   Smart meters    &    Distribution feeder devices   \\ \hline
\bfseries Applications       &    Serves the customers   &   Serves the utility    \\ \hline 
\bfseries Domain       &    Customer domain   &   Distribution domain    \\ \hline    
\end{tabular}
\end{table*}

\section{\mub{Comparison with Related Survey Articles}}
\label{comparsur}

Our current survey article is unique in a sense that it comprehensively covers the area of SDN-based \drmub{SGC}. There is no prior detailed survey article that jointly considers \drmub{SDNs} and \drmub{SG}, to the best of our knowledge. Though there is an extensive literature on survey articles on \drmub{SDNs} or \drmub{SG}, but these survey articles either focus on \drmub{SDNs} or \drmub{SG}, individually. 

General survey articles covering the broader picture of \drmub{SG} are discussed in~\cite{FaMX14,LoAn12,MoGR14}. Survey articles dealing with communication aspects of \drmub{SG} are presented in~\cite{ArendP75,ArendP19,ArendP05,ArendP87,sau2011end,ArendP11,ArendP20,gun2011sma,ArendP66,Khan16comst,Khan17commag,Rehmani16adhocnet}. Routing schemes and networking issues in \drmub{SG} have been discussed in~\cite{sab2014sur,ArendP10}. Demand response (DR) for \drmub{SG} is specifically discussed in~\cite{den2015sur,VaZV14,ArendP90}. Studies on energy efficiency for SG have been surveyed in~\cite{ArendP91,KaMoCOMST14}, \drmub{while security and privacy related work for SG are surveyed in~\cite{YaQS12,LiXL12,ArendP92,Habib18Tia,Stellios18comst}}. The discussion in~\cite{ArendP89} covers load balancing in \drmub{SG}. Simulations to support power and communication network for \drmub{SG} system analysis is presented in~\cite{ArendP88}. Use of wireless sensor networks in \drmub{SG} is surveyed in~\cite{fad2015sur,gun2010opp}. Stochastic information management in SG is discussed in~\cite{ArendP86}, while context awareness is presented in~\cite{don2015con}. SG neighborhood area networks (NANs) are discussed in~\cite{ArendP12}. 

From the perspective of \drmub{SDNs}, one can find numerous survey articles. General discussion on SDN technology is presented in~\cite{Farhady15comnet,Rahim16jnca,Nunes14comst}. Fault management schemes for SDN have been reviewed in~\cite{Fonseca17comst}, while traffic engineering through SDN is surveyed in~\cite{Mendiola17comst}. Transport network, topology discovery, and routing for SDN are surveyed in~\cite{Alvizu17comst},~\cite{Khan17comst}, and~\cite{Guck18comst}, respectively. The discussion in~\cite{Mijumbi16comst} covers the comparison of network function virtualization (NFV) and SDN technologies. Security issues for \mub{SDN are discussed} in~\cite{Rawat17comst,Dargahi17comst,Scott16comst}. A survey article on Hypervisors (which isolates the underlying physical \drmub{SDN} and its devices into virtual SDN network), has been presented in~\cite{Blenk16comst}. A survey on testbed for SDN is provided in~\cite{Huang17comst}.  

In the literature, we can find few articles on the use of SDN in SG~\cite{Zhang13cicsp,Dong15cpss,Leal16lat,Dorsch14smartgridcom,Whou17commag,Jaebeom15aps}. However, these articles are not comprehensive. For instance, \drmub{in~\cite{Zhang13cicsp}, the authors discussed a few} opportunities for the use of SDN in \drmub{SG}, covering only nineteen articles published back till 2013. Authors in~\cite{Dong15cpss} only \drmub{discussed} how SDN can be used to increase the resilience of SG against malicious attacks. On top of it,~\cite{Leal16lat} is in Spanish, thus, hampered the general English readers to read and understand it. One closely related article is~\cite{Dorsch14smartgridcom}, however, this article is in fact not a survey article and merely discussed few use cases and a testbed is introduced. Very recently, an article is published~\cite{Whou17commag}, but the focus in on machine-to-machine (M2M) communication in SDN based smart energy management. Another recent article is published\mub{~\cite{Molina18caee}}, however, it focus on cyber physical systems in general and have not provided in-depth discussion of applying SDN into SG systems. Thus, our article presents an up-to-date comprehensive review of \drmub{SDN} based \drmub{SG}, including advantages, architectures, routing, and security schemes. We also outline open issues, challenges, and future research directions related to \drmub{SDN-based SGC}. \mub{We provide the summary comparison of survey article on SDN and SG in Table~\ref{tab:surveysgsdn}. A detailed comparison of survey articles closely related to \drmub{SDN-based SGC} is presented in Table~\ref{tab:sursdnbasedsg}.}

\begin{table*}[t]\footnotesize
\centering
\caption{\mub{Comparison of Survey Articles on \drmub{SG} and \drmub{SDNs}}}
\label{tab:surveysgsdn}
\begin{tabular}{|p{3.2cm}|p{8cm}|p{2cm}|p{2.2cm}|}
\hline
\bfseries \mub{Main Domain} & \bfseries \mub{Sub-Topic} &\bfseries \mub{Reference} &\bfseries \mub{Publication Year} \\
\hline
\multirow{31}{*}{} &  & \multirow{3}{*}{} \mub{~\cite{FaMX14}} & \mub{2012} \\ \cline{3-3} \cline{4-4}  
                  & \mub{General Introduction of SG} &  \mub{~\cite{LoAn12}}  & \mub{2012} \\ \cline{3-3} \cline{4-4}  
                  &  &  \mub{~\cite{MoGR14}}  & \mub{2014} \\ \cline{2-4} 
                  & \multirow{12}{*}{} &  \mub{~\cite{Khan17commag}}   & \mub{2017} \\ \cline{3-3} \cline{4-4}
                  &  &  \mub{~\cite{gun2011sma}}   & \mub{2011} \\ \cline{3-3} \cline{4-4} 
                  &  &  \mub{~\cite{ArendP11}}   & \mub{2013} \\ \cline{3-3} \cline{4-4} 
                  &  &  \mub{~\cite{ArendP20}}   & \mub{2013} \\ \cline{3-3} \cline{4-4} 
                  &  &  \mub{~\cite{ArendP75}}  & \mub{2012} \\ \cline{3-3} \cline{4-4} 
                  & \mub{Communication Aspects} &  \mub{~\cite{ArendP19}}    &  \mub{2013}\\ \cline{3-3} \cline{4-4} 
                  &  &  \mub{~\cite{ArendP05}}    & \mub{2011} \\ \cline{3-3} \cline{4-4} 
                  &  &  \mub{~\cite{ArendP87}}   & \mub{2014} \\ \cline{3-3} \cline{4-4} 
                  &  &  \mub{~\cite{sau2011end}}      & \mub{2011} \\ \cline{3-3} \cline{4-4} 
                  &  &  \mub{~\cite{ArendP66}}   & \mub{2013} \\ \cline{3-3} \cline{4-4}
                  &  &  \mub{~\cite{Khan16comst}}  & \mub{2016} \\ \cline{3-3} \cline{4-4}  
                  &  &  \mub{~\cite{Rehmani16adhocnet}}   & \mub{2016} \\ \cline{2-4} 
                  & \multirow{2}{*}{}\mub{Routing and Networking Issues} & \mub{~\cite{sab2014sur}}  & \mub{2014} \\ \cline{3-3} \cline{4-4}  
                  &  & \mub{~\cite{ArendP10}}  & \mub{2014} \\ \cline{2-4}                           
\mub{Smart Grid}        & \multirow{3}{*}{} &  \mub{~\cite{den2015sur}} & \mub{2015} \\  \cline{3-3} \cline{4-4}   
                  & \mub{Demand Response} &  \mub{~\cite{VaZV14}} & \mub{2015} \\  \cline{3-3} \cline{4-4} 
                  &  & \mub{~\cite{ArendP90}}  & \mub{2014} \\ \cline{2-4} 
                  & \multirow{2}{*}{}\mub{Energy Efficiency} & \mub{~\cite{ArendP91}}  & \mub{2014} \\ \cline{3-3} \cline{4-4} 
                  &  & \mub{~\cite{KaMoCOMST14}}  & \mub{2015} \\ \cline{2-4}  
                  & \multirow{5}{*}{}  & \mub{~\cite{YaQS12}} & \mub{2012} \\  \cline{3-3} \cline{4-4}
                  & \mub{Security and Privacy} & \mub{~\cite{LiXL12}}  & \mub{2012} \\ \cline{3-3} \cline{4-4} 
                  &  & \mub{~\cite{ArendP92}} & \mub{2014} \\ \cline{3-4} 
                  &  & \mub{~\cite{Habib18Tia}} & \mub{2018} \\ \cline{3-4}
                  &  & \drmub{~\cite{Stellios18comst}} & \mub{2018} \\ \cline{2-4}                                  
                  & \mub{Load Balancing} & \mub{~\cite{ArendP89}} & \mub{2014} \\ \cline{2-4} 
                  & \mub{Simulation to Support Power and Communication Network} & \mub{~\cite{ArendP88}}  & \mub{2014} \\ \cline{2-4} 
                  & \multirow{2}{*}{}\mub{Use of Wireless Sensor Networks in SG} &  \mub{~\cite{fad2015sur}}  & \mub{2015} \\ \cline{3-3} \cline{4-4} 
                  &  &  \mub{~\cite{gun2010opp}}  & \mub{2010} \\ \cline{2-4}
                  & \mub{Stochastic Information Management} &  \mub{~\cite{ArendP86}}   & \mub{2014} \\ \cline{2-4}  
                  & \mub{Context Awareness} &  \mub{~\cite{don2015con}}  & \mub{2015} \\ \cline{2-4}                                               
                  & \mub{SG NANs} &  \mub{~\cite{ArendP12}}    & \mub{2014} \\ \hline
\multirow{14}{*}{} & \multirow{2}{*}{} & \mub{~\cite{Farhady15comnet}}  & \mub{2015} \\ \cline{3-3} \cline{4-4} 
                  & \mub{General Introdcution of SDN} &  \mub{~\cite{Rahim16jnca}} & \mub{2016} \\ \cline{3-3} \cline{4-4}
                  &  &  \mub{~\cite{Nunes14comst}} & \mub{2014} \\ \cline{2-4}
                  & \mub{Fault Management} &  \mub{~\cite{Fonseca17comst}} & \mub{2017} \\ \cline{2-4}
                  & \mub{Traffic Engineering} & \mub{~\cite{Mendiola17comst}}  & \mub{2017} \\ \cline{2-4}
                  & \mub{Transport Network} & \mub{~\cite{Alvizu17comst}} & \mub{2017} \\ \cline{2-4}
                  & \mub{Topology Discovery} & \mub{~\cite{Khan17comst}} & \mub{2017} \\ \cline{2-4}
\mub{Software Defined Networks}        & \mub{Routing} &  \mub{~\cite{Guck18comst}} & \mub{2018} \\ \cline{2-4}
                  & \mub{Comparison of NFV and SDN} &  \mub{~\cite{Mijumbi16comst}}  & \mub{2016} \\ \cline{2-4}
                  & \multirow{3}{*}{} & \mub{~\cite{Rawat17comst}}  & \mub{2017} \\ \cline{3-3} \cline{4-4}
                  & \mub{Security} &  \mub{~\cite{Dargahi17comst}} & \mub{2017} \\ \cline{3-3} \cline{4-4}
                  &  & \mub{~\cite{Scott16comst}} & \mub{2016} \\ \cline{2-4}
                  & \mub{Hypervisors} &  \mub{~\cite{Blenk16comst}} & \mub{2016} \\ \cline{2-4}
                  & \mub{Testbed for SDN} &  \mub{~\cite{Huang17comst}}  & \mub{2017} \\ \hline
\end{tabular}
\end{table*}

\begin{table*}[t]\footnotesize
\centering
\caption{\mub{Comparison of Survey Articles on SDN-based \drmub{SGC}. \cmark Indicates that the topic is covered, \xmark indicates that the topic is not covered, and \ding{93} indicates that the topic is partially covered.}}
\label{tab:sursdnbasedsg}
\begin{tabular}{|p{1.1cm}|p{1.2cm}|p{2.4cm}|p{1cm}|p{1cm}|p{1.4cm}|p{1.4cm}|p{1cm}|p{1.4cm}|p{1.4cm}|}
\hline
\bfseries \scriptsize \mub{Reference} & \scriptsize \bfseries \mub{Publication Year} & \scriptsize \bfseries \mub{Main Domain} & \bfseries \scriptsize \mub{Resilience} & \bfseries \scriptsize \mub{Scalability} & \bfseries \scriptsize \mub{Traffic Optimization} & \bfseries \scriptsize \mub{Architectures} & \bfseries \scriptsize \mub{Routing} & \bfseries \scriptsize \mub{Security and Privacy} & \bfseries \scriptsize \mub{Future Research Directions} \\ 
\hline
\mub{~\cite{Dorsch14smartgridcom}} & \mub{2014} & \drmub{SDN-based SGC} & \mub{\cmark} & \mub{\xmark} & \mub{\cmark} & \mub{\cmark} & \mub{\cmark} & \mub{\xmark} & \mub{\xmark}   \\ \cline{1-10} 
\mub{~\cite{Zhang13cicsp}} & \mub{2013} & \drmub{SDN-based SGC} & \mub{\xmark} & \mub{\xmark} & \mub{\xmark} & \mub{\cmark} & \mub{\xmark} & \mub{\xmark} & \mub{\xmark}   \\ \cline{1-10} 
\mub{~\cite{Dong15cpss}} & \mub{2015} & \drmub{SDN-based SGC} & \mub{\cmark} & \mub{\xmark} & \mub{\xmark} & \mub{\xmark} & \mub{\xmark} & \mub{\cmark} & \mub{\xmark}   \\ \cline{1-10} 
\mub{~\cite{Leal16lat}} & \mub{2016} & \drmub{SDN-based SGC} & \mub{\xmark} & \mub{\xmark} & \mub{\xmark} & \mub{\cmark} & \mub{\xmark} & \mub{\xmark} & \mub{\xmark}   \\ \cline{1-10} 
\mub{~\cite{Whou17commag}} & \mub{2017} & \mub{SDN-based M2M} & \mub{\xmark} & \mub{\cmark} & \mub{\cmark} & \mub{\cmark} & \mub{\xmark} & \mub{\cmark} & \mub{\ding{93}}   \\ \cline{1-10} 
\mub{~\cite{Jaebeom15aps}} &  \mub{2015} & \drmub{SDN-based SGC} & \mub{\xmark} & \mub{\xmark} & \mub{\cmark} & \mub{\cmark} & \mub{\cmark} & \mub{\cmark} & \mub{\ding{93}}   \\ \cline{1-10} 
\mub{~\cite{Molina18caee}} &  \mub{2018} & \mub{SDN-based CPS} & \mub{\cmark} & \mub{\cmark} & \mub{\xmark} & \mub{\xmark} & \mub{\xmark} & \mub{\xmark} & \mub{\ding{93}}   \\ \cline{1-10} 
\scriptsize \mub{This Work} & \mub{2018} & \drmub{SDN-based SGC} & \mub{\cmark} & \mub{\cmark} & \mub{\cmark} & \mub{\cmark} & \mub{\cmark} & \mub{\cmark} & \mub{\cmark}   \\ \cline{1-10} 
\hline
\end{tabular}
\end{table*}

\section{\drmub{SG} and \drmub{SDNs}: Background, Terminology, and Definitions}
\label{sgbk}

\drmub{This section briefly reviews the background, terminology, and definitions related with SG and SDN. Then we discuss the issues specific to SGC. Finally, we discuss case studies on the use of SDN-based SGC. }

\subsection{Smart Grid}
\label{sg}

The traditional electric power grid mainly supports four operations: electric power generation, electric power transmission, electric power distribution, and the control of generated electricity. With the passage of time, this electric power grid was getting older and was unable to support new power services and applications. Moreover, due to centralized generation, old infrastructure, lack of control, and one-way communication, failures and blackouts were getting frequent. In order to address these challenges and to consider the future electric needs, \drmub{information and communication technologies (ICT)} with advanced control strategies, seems to be the indispensable part of the future electric grid, also known as the future \drmub{SG}. The eventual goal of \drmub{SG} is to maintain reliable electric supply, to accommodate distributed \drmub{RERs}, to reduce greenhouse gas emission, and to automate the operations and maintenance of the electric power grid~\cite{FaMX14}.   

From the architectural point of view, \drmub{SG} is composed of three building blocks: Home Area Networks (HANs), Neighborhood Area Network (NANs), and Wide Area Networks (WANs). The connectivity of distributed \drmub{RERs}, Plug-in Electric Vehicles (PEVs), consumer devices, and smart meters present within the premises of consumers will be the responsibility of HANs. \mub{HANs are thus responsible for the charging of PEVs. In addition, it is estimated that homes are the places where 50\% of total electricity is consumed~\cite{Heile10wcm}.} Multiple HANs will be connected through the NANs. NANs are composed of \drmub{AMI} and field area networks (FANs). A comparison of AMI and FANs is elaborated in Table~\ref{compamifans}~\cite{Chang13wc}. WANs is responsible to connects NANs with power utility facilities and control center. 

In order to control and manage the \drmub{SG}, and to support diverse emerging consumer-side and utility-side applications, enhanced communication technologies are necessary~\cite{Gungor13tii}. These communication technologies can be wired or wireless, depending upon the utility and the application needs. A detailed description of these \drmub{ICT} and infrastructure to support SG operation can be found in~\cite{KaMoCOMST14}. 

\mub{Despite all these advantages that SG will bring, there are certain issues which are specific to SG communication system. We now provide a brief overview of some of these issues. }

\subsubsection{\drmub{Issues Specific to SGC}}

\begin{itemize}
\item {\drmub{\it \bfseries Resilient to Attacks and Failures:} SG systems are not too much resilient to attacks and failures. In case of any communication link failure or an attack, the SG system should be capable to respond quickly and restore its operational state. In this context, the SG communication system and devices need to be easily programmable so that they quickly adapt to the changing conditions occurring at the network level. }
\item {\drmub{\it \bfseries Vendor Specific Devices and Protocols:} \mub{SG communication system is built on different vendor specific devices and protocols. Therefore, the current SG systems are not protocol independent, thus leading to interoperability issue. This interoperability issue hinders SG to implement and run diverse applications in large variety of networking technologies and protocols. }}
\item {\drmub{\it \bfseries Granularity:} \mub{Granularity is another issue which arises due to different vendor specific devices and protocols. More precisely, the switches designed by one vendor may offers traffic monitoring at flow level, while the switches designed by other vendor may provide traffic identification at packet level. This will lead to synchronization problem in traffic and flow management for the SG communication system. }}
\item {\drmub{\it \bfseries Security and Privacy:} \mub{Security and privacy are also the main issues which current SG systems are facing (see Section~\ref{secur} for more details). For instance, there are particular types of attacks (link flood attacks, and smart meter data manipulation attacks) which are specifically designed to deteriorate the performance of SG system.}}
\item {\drmub{\it \bfseries Network Management:} \mub{Network management of current SG systems is complex, time consuming, and done manually. This also includes manual intervention from the network administrators and network engineers to restore the operational state of the SG system. To illustrate this more, we present here an example of a campus-based microgrid which has established in the British Columbia Institute of
Technology (BCIT). It consists of three networks namely HAN, LAN, and the WAN. To make LAN operational, ZigBee networks has adopted. Since ZigBee networks receive severe interference from
\drmub{Wi-Fi} networks, therefore, it was concluded that channel 20 should be used to avoid interference with the \drmub{Wi-Fi} network in residential area. However, this selection of channel 20 need to be managed by the SG network engineers manually. This is just one example which illustrate that even simple SG network (in this case LAN based on ZigBee) has to be carefully managed~\cite{Stanciulescu12isgt}. In Section~\ref{adv}, we discuss how \mub{SDN-based SGC} will address these aforementioned issues.}}
\end{itemize}

\subsection{Software Defined Networks}
\label{sdn}
The idea of \drmub{SDN} has emerged from the need for programmable networks. Traditional networks, such as Internet, are not much programmable. The building block of Internet are the devices such as switches, and routers which need to be configured by the network operators. Network operators implement policies on these devices so that these devices respond to network events and particular applications. However, the configuration and implementation of policies are done manually and not flexible enough to interact with the dynamic environment of the Internet and new emerging applications. Here SDN with its ``programmability'' feature separates the hardware from the control decisions. In simple words, network devices, such as switches or routers, become forwarding devices and the software defined controllers led the network intelligence~\cite{Nunes14comst}. SDN has been proposed to monitor and manage the communication networks globally. The applicability of SDN in different domains is not new. SDN has been applied to data centers \cite{Singh17jpdc,Cui13jsac}, \drmub{WAN} \cite{Ahmed14commag}, enterprises \cite{Chen16tetc,Lorenz17commag}, optical networks \cite{Thya16comst}, wireless networks \cite{Haque16comst}, wireless sensor networks \cite{Kobo17access}, and under water sensor networks (UWSN) \cite{Akyildiz16adhocnet}.

\begin {figure*}
\centering
\begin{adjustbox}{width=\linewidth}
\tikzstyle{block} = [rectangle, draw, text width=3.5cm, text centered, rounded corners, minimum height=3em,fill=blue!20]
\tikzstyle{line} = [draw,thick, -latex']
\tikzstyle{cloud} = [draw, ellipse, text width=2.5cm, text centered]
\tikzstyle{edge from parent}=[->,thick,draw]
\begin{tikzpicture}[auto,edge from parent fork down]
\tikzstyle{level 1}=[sibling distance=70mm,level distance=16ex]
\tikzstyle{level 2}=[sibling distance=40mm,level distance=24ex]
\tikzstyle{level 3}=[sibling distance=35mm,level distance=34ex]
\node [block,text width=8cm,minimum height=3em,,fill=red!40] (cst) {Case Studies on the Use of \drmub{SDN-based SGC} \\ Sect. \ref{casestu}}
{
child{node [block,fill=green!40] (pmt) {\footnotesize Substation Automation and Monitoring \\ Sec. \ref{submon}\\ \footnotesize \cite{Cahn13smartgridcom} \mub{(2013)},
\cite{Leal16latincom} \mub{(2016)}, \cite{Meloni17energies} \mub{(2017)}}}
child{node [block,fill=green!40] (opm) {\footnotesize  Utility M2M Applications \\ Sec. \ref{um2m}\\ \footnotesize \cite{Whou17commag} \mub{(2017)}, \cite{Kim14smartgridcom} \mub{(2014)}, \cite{Pozuelo16icsgs} \mub{(2016)} }}
child{node [block,fill=green!40] (rnm) {\footnotesize  Cloud and IoT based Applications \\ Sec. \ref{cloiot}\\ \footnotesize \cite{Xin11smartgridcom} \mub{(2011)}  }}
};
\end{tikzpicture}
\end{adjustbox}
\caption {Case studies on the use of \drmub{SDN-based SGC} in the context of different scenarios.}
\label{fig:casestu}
\end{figure*}
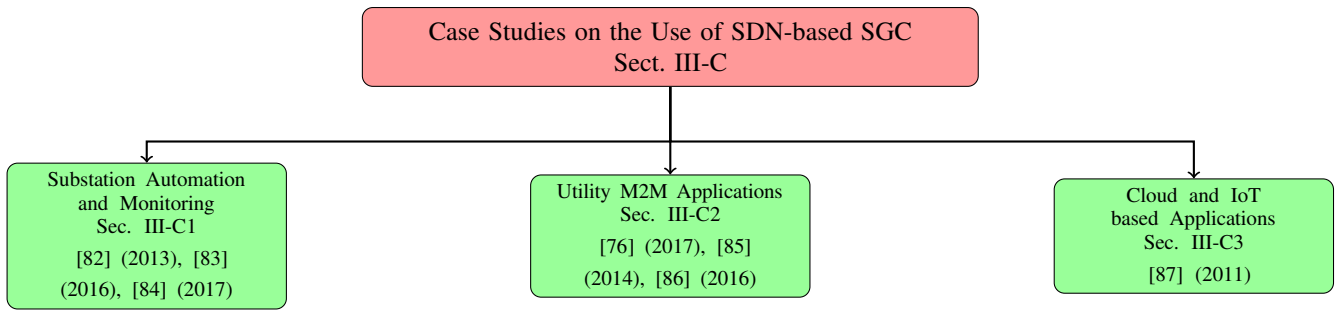

\drmub{The key principle of SDN is to provide logically centralized control through SDN centralized controller~\cite{Jarschel14commag}. SDN's centralized control architecture enables it to maintain a global view of the network states and information. As a consequence, SDN by using its centralized controller can take forwarding and routing decisions much faster and better. For instance, consider the current Internet where traditional system of routers  can exchange link state information using a certain protocol, then update forwarding tables in a distributed manner. Such a distributed control with proprietary middle boxes and routers makes it virtually impossible to adopt a new routing protocol (e.g., IP multicast) because of the requirement of global deployment. However, in SDN, the SDN controller determines forwarding table for each switch. In such a way, a new routing scheme, such as IP multicast, can be flexibly supported. However, one may argue that SDN centralized control will not be scalable with respect to network size. Though from the perspective of scalability, realizing logically centralized controller is challenging but depending upon the network size, SDN controller can consists of a distributed system of multiple SDN controllers of physical and virtual instances (i.e., distributed control plane) behaving logically as a single entity~\cite{Bannour18comst}.}

SDN leverages multitude of functionalities and access to fine grained packet related information through SDN controllers, such as OpenFlow and OpenDayLight~\cite{Hansen15commag}. This information help the SDN controllers to access packet collision related information, port information, hardware description, and the type of connection used. Moreover, SDN controllers are capable to dynamically configure the flow entries on switches and routers~\cite{Jarschel14commag}. On top of it, SDN controllers can also identify errors in data paths, unidentified packets, and may remove or alter the data flow path entries. All these capabilities make the SDN controller very powerful and thus highly suitable for SG communications. 

Traditional approaches, like \drmub{MPLS}, has been adopted initially by the utilities for \drmub{SG} communication system but it is not completely sufficient. One primary reason for not adopting MPLS in SG is because with the addition of new services in SG, the MPLS based routers need to be re-configured each time, resulting in disruption of services provided by the utilities \cite{Sydney13tsg}. Distributed software have been proposed for SG but cannot replace the effectiveness of SDN \cite{Patti16tsg}. Thus, one alternative that comes in our mind is \drmub{SDN}, which further led to the emergence of \drmub{SDN} based \drmub{SG} communication (SDN-based SGC). 

\subsubsection{OpenFlow Protocol} OpenFlow~\cite{Openflow} is the most popular standard/protocol for SDN and proposed by Stanford~\cite{Hu14comst}. The OpenFlow architecture consists of switches, controllers and flow entries. The OpenFlow switch contains channel, group table, and flow tables. The OpenFlow switch communicates with the SDN controller through OpenFlow protocol. Moreover, OpenFlow has been widely adopted by the industry, for instance, it has been used by data centers, mobile applications, and for industry research as well. OpenFlow protocol is used by the SDN controller (OpenFlow controller). The OpenFlow controller is responsible to install flow rules in the flow tables of the OpenFlow switches. Whenever a packet comes to the OpenFlow switch, it check its flow table and the corresponding action is taken. If an entry does not exist in the flow table of the OpenFlow switch, the OpenFlow switch forward it to the OpenFlow controller. The OpenFlow controller then process this packet for this particular type of packet and new flow rules are added to the OpenFlow switches. Then in future, if new such packet type comes, the OpenFlow switch see flow table entry and process the packet without consulting the OpenFlow controller again.

\begin{figure*}[t] \centering
\includegraphics[width=0.9\textwidth]{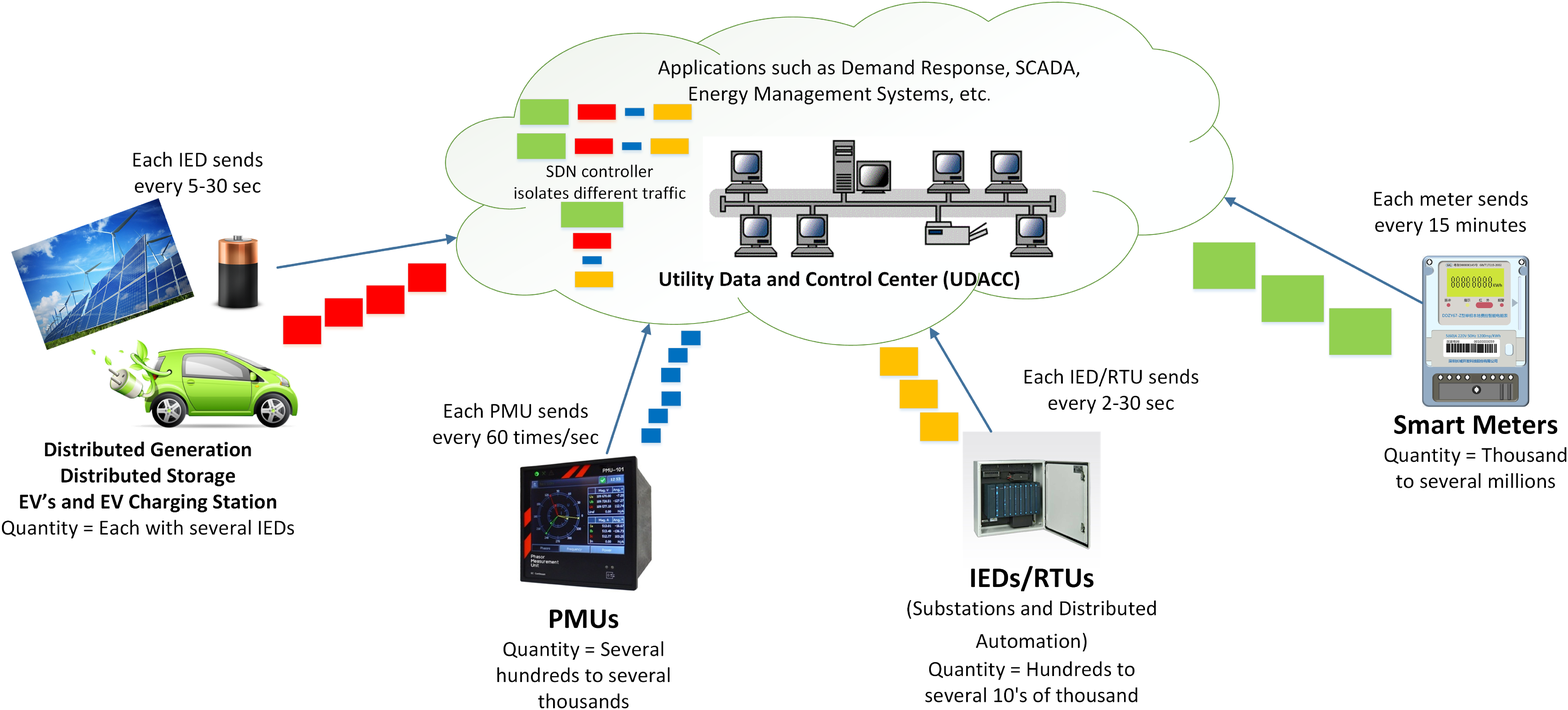}
\caption{\mub{Utility Data and Control Center is receiving different types of traffic generated by different devices at different rates. This traffic can be event-based or generated at regular intervals. By using SDN-based SGC (UDACC in this case), different traffic types and/or applications can be easily isolated.}} 
\label{udacc}
\end{figure*}

\subsection{\mub{Case Studies on the Use of \drmub{SDN-based SGC}}}
\label{casestu}

In this section, we briefly discuss three case studies on the use of \drmub{SDN-based SGC} systems applied to different scenarios. We also illustrate these case studies in Fig.~\ref{fig:casestu}.

\subsubsection{Substation Automation and Monitoring}
\label{submon}
This case study shows how \drmub{SDN-based SGC} can be used in substation automation and monitoring. A lot of work has been done for substation automation and monitoring~\cite{Cahn13smartgridcom,Leal16latincom,Meloni17energies,Gopal16icps}.

In a substation environment, researchers and manufacturers have been struggling to automate the functionality of substations. In this context, SDN is the best candidate technology as it provides liberty to the utilities and substation administrator to automate the communication and actuation tasks as much as possible. \mub{In this context, Leal et al.~\cite{Leal16latincom} proposed a SDN based communication architecture for SG automation.} The proposed architecture is called as smart solution for substations networks (S3N). S3N builds on three layers: Infrastructure layer, virtualization layer, and functionality layer. The infrastructure layer is responsible for the management of physical devices and resources. The virtualization layer helps to provide virtual resources, and the functionality layer supports various functionalities. S3N consists of four modules: The first module provide protect and control to the substation (S3N-PROTECT), the second module provides management features to substation (S3N-MANAGE), the third module provides measurement capability in substation (S3N-MEASURE), and finally the fourth module provides connectivity among different substation equipments (S3N-CONNECT). In summary, the S3N architecture provides framework for the automation of substation.

\mub{For auto-configuration of substation in SG, Cahn et al.~\cite{Cahn13smartgridcom} proposed a software defined energy communication network (SDECN) architecture.} The SDECN architecture will help substations to meet with future requirements of the SG. For evaluation purpose, authors used RYU based SDN controller and developed its prototype. One of the unique feature of SDECN is that it gives the liberty to \mub{not configure multiple VLANs} for traffic isolation and this traffic isolation can be easily done through SDECN. Authors evaluated SDECN through Mininet emulations and demonstrated that unlike traditional network containing typical switches, there is no need to configure the substation communication network. 

Another related study~\cite{Gopal16icps} provides uses cases for the automation of \drmub{SG}. Issues such as auto-configuration of plug-and-play of IEDs and smart appliances, registration of RERs, and long term planning were discussed in detail.  

\subsubsection{Utility M2M Applications}
\label{um2m}
This case study shows how \drmub{SDN-based SGC} can be used in utility M2M applications. There are few works reported on SDN-based utility M2M applications~\cite{Pozuelo16icsgs,Whou17commag,Kim14smartgridcom}. \drmub{In a Utility M2M scenario, different types of electronic devices such as smart meters, IEDs, PMUs, and sensors are connected with the Utility Data and Control Center (UDACC) and communicate with each other. Thanks to the advancement in field programmable gate array (FPGA), software defined meters is now in the market for utility M2M applications~\cite{Depari16csi}. These meters are the building blocks for smart utility networks (SUNs). Fig.~\ref{udacc} shows Utility M2M applications along with their communication frequency. As can be seen in Fig.~\ref{udacc}, UDACC is receiving different types of traffic generated by different devices at different rates. This traffic can be event-based or generated at regular intervals. By using SDN in SGC (UDACC in this case), different traffic types and/or applications can be easily isolated (cf. Fig.~\ref{udacc})~\cite{Kim14smartgridcom}.}

Similarly, ~\cite{Whou17commag} provides a case study of SDN-M2M. Authors considered an electric vehicle energy management (EVEM) system based on SDN. In the presented SDN-based EVEM, authors considered 100 EVs with one gas generator and four wind turbines. The SDN controller is responsible for keeping the information of EVs such as their battery charging status, location, and charging time. In this EVEM system, SDN helps to improve mobility management of EVs and helps to perform resource allocation easily. To elaborate it further, let's consider a scenario in which M2M devices (EVs) increases and thus collision probability increase due to random access of EVs. Thus, critical information cannot be delivered to EVs. If SDN is employed, the SDN controller can generate resource allocation block depending upon specific QoS requirements, thus reducing the number of competing M2M devices. This will ultimately reduce collision probability and therefore critical information can be timely communicated.

\subsubsection{Cloud and IoT based Applications}
\label{cloiot}
The case study~\cite{Xin11smartgridcom} shows how \drmub{SDN-based SGC} can be used in conjunction with cloud and Internet of Things (IoT). \mub{Towards this goal, Xin et al.~\cite{Xin11smartgridcom} proposes a virtual SG architecture to embed with the cloud in which infrastructure as a service (IaaS) is used by virtualizing the SG hardware.} The focus of this virtual SG architecture is SG transmission applications. \\

We have discussed three case studies in this section to demonstrate the use of SDN-based \drmub{SGC} in different scenarios. These three examples clearly shows that SDN-based \drmub{SGC} is a viable solution to adopt in future to support future \drmub{SG} communication management.

\begin{table*}[t]\footnotesize
\centering
\caption{Summary comparison of \drmub{SG} and SDN-based \drmub{SGC}}
\label{tab:compsgsdnsg}
\begin{tabular}{|p{2.5cm}|p{7cm}|p{7cm}|}
\hline
\bfseries Parameter & \bfseries Smart Grids &\bfseries SDN-based Smart Grid  \\
\hline
Programmability & SGs are not highly programmable	& With SDN capability, SGs now are easily programmable \\
 \hline
Protocol independence  & Not truly protocol independent	& Protocol independence can be easily achieved through SDN controllers \\
 \hline
Granularity  & Dependent upon proprietary hardware	& SDN controllers can identify the traffic at every flow and packet level \\
 \hline
Resilience & Not too much resilient to attacks and failures	& SG resilience against failures and malicious attacks can be achieved by using SDN \\
 \hline
Network Management & Complex, time consuming, and Manual & Easy, automatic, and faster \\
\hline
Interoperability & Difficult to cope with different vendor specific devices and protocols & SDN technology is not vendor specific and operates on open standards. Thus, various types of
communication network devices can be easily managed and configured and their interoperability will not be a
problem within a SG \\
\hline
Simulation Tools \& Testbeds  & Lot of them are available	& Need to develop \drmub{SDN-based SGC} simulation tools and testbeds \\
 \hline
Standardization  & Low of work is done on SG standardization	& Need more efforts for SDN specific standards for SG \\
 \hline
Security and Privacy  & Several security and privacy schemes are proposed	&  May need to develop new algorithms as SDN controller may compromised or SDN controller applications may get compromised \\
 \hline
\end{tabular}
\end{table*}

\section{Taxonomy of Advantages of \drmub{SDN-based SGC}}
\label{adv}

\drmub{Based on the discussion of SG, SDN, and use cases of \drmub{SDN-based SGC} introduced in Section~\ref{sgbk}, we are now ready to review the motivations for adopting SDN in the SG. In this section, we start by providing the general motivations for adopting SDN in the SG. Then we discuss the taxonomy of advantages of \drmub{SDN-based SGC} in the areas of SG resilience, SG stability, SG traffic optimization and other advantages.}

\subsection{Motivations for Adopting SDN in the SG}
\label{motiv}

Table~\ref{tab:compsgsdnsg} summarizes the comparison of traditional \drmub{SG} and SDN-based \drmub{SGC}. \drmub{SG} generates different types of traffic and these traffic types have diverse Quality of Service (QoS) requirements in terms of reliability, delay, and throughput \cite{Khan17commag}. \mub{For instance, wide area situational awareness traffic in SG requires 99.99\% reliability and 20-200 ms delay, while real-time pricing from utility to meters require greater than 98\% reliability and less than 1 min delay~\cite{Khan17commag}.} In such situations, the SDN controller can identify the traffic type and then prioritize the traffic by programming dynamically the SDN enabled switches in SG environment. Another scenario is the bulk transfer of meter readings from \mub{AMI} to the utility. \mub{This traffic type} is not delay sensitive and may require transferring \mub{Megabytes} of data. Thus, the SDN controller can prioritize the traffic to increase the throughput and to decrease the number of transmissions while transmitting it from AMI to utility.

The SDN controller, by using the {\it programmability feature}, will decide in which situation, it has to use a certain link and this can be decided based upon variations in SG communication traffic. Different SG components follow different standards and protocols \cite{gun2011sma,Kim17commag,ArendP11,ArendP20} and SDN controller should be able to cope with all such diverse communication systems. The {\it protocol independence feature} of SDN will help SG to meet with this interoperability issue and help SG to implement and run diverse applications in large variety of networking technologies and protocols. The {\it granularity feature} of SDN will help SG to perform traffic flow orchestration \cite{Dorsch14smartgridcom}, to manage traffic prioritization, and to meet QoS specific requirements.

In the following, we summarize few main motivations for employing SDN in the SG:

\begin{itemize}
\item {\it \bfseries Isolation of Different Traffic Types/Applications}: In \drmub{SG}, different types of traffic are generated by different devices. This traffic can be event-based or generated at regular intervals. By using SDN in SG, different traffic types and/or applications can be easily isolated~\cite{Kim14smartgridcom}. Moreover, \drmub{SDN-based SGC} can adapt PMU's measurement data traffic according to the capabilities of receiving devices (see Section~\ref{sttraf}).
\item {\it \bfseries Traffic Prioritization:} In \drmub{SG} environment, critical measurement data and control commands need to be delivered in a timely basis and require high priority than the normal traffic. SDN can help in this regard by prioritizing the traffic and give highest priority to sensitive time critical control commands and measurement data in a flexible manner~\cite{Dorsch16smartgridcom}. Additionally, SDN based programmable controller have global network view. Therefore, it will help to orchestrate traffic flows easily (see Section~\ref{sttraf}).
\item {\it \bfseries Virtual Network Slices:} SDN can help to create virtual network slices in the \drmub{SG} based on geographical or domain consideration (transmission and distribution or security zones)~\cite{Kim14smartgridcom}. For instance, AMI network can create its own virtual network having its own virtual network slice. This will enable the AMI network to have its own security, management, and QoS policies. 
\item {\it \bfseries Resilience:}  \drmub{SG} resilience can be easily achieved through SDN by directing traffic flow from broken wired links to wireless link \cite{Aydeger15nfvsdn}. By doing this, SG will become more reliable. Furthermore, self-healing mechanism to achieve resilient PMU network can easily be done by using SDN~\cite{Aydeger15nfvsdn} (see Section~\ref{sgres}).
\item {\it \bfseries Fast Failure Recovery:} \drmub{SG} heavily relies on communication links. If these communication links get congested or broken down then the SG will not function properly. Therefore, link failure detection and recovery is essential. By applying SDN in SG, one can achive fast link  failure recovery~\cite{Dorsch16smartgridcom} (see Section~\ref{sgstab}).
\item {\it \bfseries Prevent Voltage Collapse and Line Overlaod:} In power system, some times it may happen that the electric grid may become overloaded and voltage collapse may occur. \mub{Timely shifting} the line load may prevent \mub{voltage collapse} in the \drmub{SG} and this can be easily achieved by using \drmub{SDN-based SGC}~\cite{Dorsch16smartgridcom} (see Section~\ref{sgstab}).
\item {\it \bfseries Interoperability:} SDN technology is not vendor specific and operates on open standards. Thus, various types of communication network devices can be easily managed and configured and their interoperability will not be a problem within a SG. 
\item {\it \bfseries EVs Integration in SG:} Electric vehicles (EVs) can be considered as moving power plants if we deploy them at a massive scale. Imagine EVs moving on highways carrying battery stored energy. The energy stored in these EVs can help the SG to balance the energy needs of the cities in emergency situation. However, the design of dynamic energy management system is required which easily update the state of these EVs, as EVs are generally mobile. However, mobility of electric vehicles and their status update will generate lot of data. In addition, joining and re-joining decisions of EVs will need reconfiguration in the \drmub{SG} system. With the help of SDN, the management complexity can be reduced substaintially~\cite{Zhang13cicsp}.
\item {\it \bfseries SDN's Run Time Configurability:} With the help of SDN's run time configurability, the QoS of the SG communication network can be improved significantly. 
\item {\it \bfseries Network Management Become Easier:} By incorporating SDN in SG, network management become easier~\cite{Zhang13cicsp}. The controller in the SG will have global view of the SG communication network and based upon the application requirement and underlying network traffic condition, the SDN controller will change the rules of processing packets quickly and easily in the switches. Without SDN, such network management and changes require manual intervention by the utilities to re-program the switches to change their packet forwarding rules. 
\end{itemize}

\drmub{The left four branches of Fig.~\ref{fadvsdnsg} shows} the taxonomy of advantages of \drmub{SDN-based SGC}. We classify these advantages into categories describing what SDN brings to the SG in terms of resilience, stability, and traffic \mub{optimization}. We now discuss each of them in detail.

\subsection{SG Resilience}
\label{sgres}

\subsubsection{\drmub{SG} Resilience}
\drmub{SG} resilience means the ability of the \drmub{SG} system to react with sudden failures and malicious attacks and in response to these failures and attacks, the \drmub{SG} should recover and maintain its critical services~\cite{Dong15cpss}. In the context of \drmub{SDN-based SGC}, \mub{a lot of work} has been done on SG Resilience~\cite{Jin17tsg,Dong15cpss,Aydeger16icc,Maziku17icnc,Lin18tsg,Zhqng16icsn,Aydeger15nfvsdn,Rubaye17iot,Ren17tsg,Ghosh16cypss,Genge16ifip}.

\mub{Aydeger et al.~\cite{Aydeger15nfvsdn} focuses on bringing resilience to SG in the context of communication link failure.} In a SG environment, if a wired link is failed then SDN controller will automatically make the wireless link up. In this manner, resilience is achieved by directing the flows from a broken wired link to a healthy wireless link. Moreover, traffic is also monitored through SDN controller. \mub{Aydeger et al.~\cite{Aydeger15nfvsdn} presented a demo in which NS-3 with Mininet is used for the evaluation purpose.} OpenDayLight is chosen as SDN controller. Manufacturing message specification (MMS) is used as a traffic to transfer. In the demo, the authors deliberately \drmub{dropped} the wired link and showed that the SDN controller updates its flow table and thus the switch will use the wireless link as a backup path to achieve resilience in SG. The main contribution in this demo is to connect NS-3 simulator with Mininet emulator to evaluate the proposed link failure scenario.

IEC 61850 is the standard designed for substation automation and control. In a \drmub{SG}, two or more substations will be joined together and thus require inter-substation communication. \mub{To address this issue, Aydeger et al.~\cite{Aydeger16icc} proposed an SDN based inter-substation communication network.} In this paper, two types of SDN controllers are introduced: Global SDN controllers and Local SDN controllers. The global SDN controller will be deployed at the central location in the utility where all the substations can be controlled. This global SDN controller will also manage the traffic flows among different substations by using IEC 61850 MMS traffic. In order to control the traffic within the substation, local SDN controllers are deployed within each substation. It is demonstrated through Mininet and NS-3 based simulations that SDN controllers help to select the wireless link with negligible delays in case of any link failure. Therefore, with the help of these SDN controllers, SG resilience is achieved by selecting redundant links when required.

\subsubsection{Self-Healing}
\label{sfheal}
\drmub{SG} self-healing means the recovery ability of the \drmub{SG} system. Wide area situational awareness and monitoring of the state of the \drmub{SG} is normally done by installing phasor measurement units (PMUs) in the field. Generally, a single PMU is deployed over a substation. PMUs are responsible for monitoring the voltage level and phasor angle of the transmission lines. This informations is collected by different PMUs and then fed into a phasor data concentrator (PDC). The PDC then forwards this collected measurement data to the control center of the utility. 

In case of a cyber attack, it may happen that a PDC or PMU or number of PMUs may get infected by the attack. If a PMU is compromised then the measurement data by that particular PMU will be lost and the system observability of the substation will no longer be available. The situation becomes worse in case the PDC gets compromised. This will result in the collapse of SG state monitoring system, as the PDC was responsible for forwarding the measurement data of several PMUs. In this case, it is clear that only the PDC is compromised but not the PMUs and they are still functional. 

To address this issue, the system adminstrator may disconnect the infected PMUs or PDC. However, this is not the optimal solution. One way to mitigate this problem is to re-route the data of PMUs to the reliable PDC. \mub{In this context, Lin et al.~\cite{Lin18tsg} provides resiliency solution to SG by proposing to establish new communication paths to deliver PMU's measurement data in order to achieve SG system observability by using SDN.} Authors \drmub{proposed} integer linear program (ILP) and a heuristic algorithm for this purpose and tested these algorithms through IEEE 30-bus and IEEE 118-bus topologies. It is demonstrated that the proposed linear program can further reduce the latency upto 75\%, while the heuristic algorithm can further reduce this latency but at the cost of incurring more overhead of upto 25\%. In this manner, self-healing is achieved in PMU network by re-routing the traffic of uncompromised PMUs and deliver their data to the PDC which is uncompromised. 

\subsubsection{Fast Failover}
Fast failover means that when a communication link failure occurs, the packets are routed to the alternative route without consulting to the SDN controller~\cite{Kumar16smartgridcom,Kurtz16netsoft,Sydney14comnet}. This feature of fast failover mechanism for SDN switches is available from OpenFlow 1.3 versions~\cite{Kumar16smartgridcom}. 

\mub{Therefore, in an effort to deal with fast failover, Kumar et al.~\cite{Kumar16smartgridcom} proposes a prototype Flow Validator which will use \drmub{North Bound Interface (NBI)} to collect information about the state of the SDN.} This information is then used by Flow Validator to incorporate fast failover mechanism to provide resiliency to SG communication. Authors showed that through Flow Validator, the SDN system will require nine times less time if changes in the link failure/restoration were computed from the scratch. 

The global environment for network innovation (GENI)~\cite{Berman15comnet} is a testbed facility provided by National Science Foundation (NSF), USA, to the researchers to evaluate thier newly developed algorithms and prototypes. GENI testbed also supports the programmability facility as SDN (OpenFlow), which is the integral part of it. Using GENI and considering the strict deadline of two weeks, the authors in~\cite{Sydney14comnet} deployed an OpenFlow SDN controller for \drmub{SG} \drmub{DR} application. In this regard, authors developed two algorithms: a control logic algorithm for load shedding and the second algorithm for link failure. The goal was to see if MPLS-like functionalities can be achieved or not in commercially available hardware switches through SDN and it was shown that such functionalities can be achieved easily. The deployed SDN controller is capable of guaranteeing the required QoS, provides fast failover mechanism, and supports load balancing.

\subsubsection{Fault Tolerance}
\label{ftol}
Fault tolerance in \drmub{SG} means the ability of the SG system to sustain operating properly even in the presence of a fault. There are lot of works done to incorporate fault tolerance capability in SG through SDN~\cite{Dorsch16globecom,Rubaye17iot,Zhqng16icsn}.

\mub{Fault tolerance mechanism for SDN based SG has been presented by Dorsch et al.~\cite{Dorsch16globecom}.} In fact, fault tolerance in SG is achieved by using SDN technology. More precisely, authors focused on link failures and identified three critical phases during link failure. The first one is link failure detection, the second one is link failure recovery, and finally the third one is post recovery optimization. 

In link failure detection, the authors of~\cite{Dorsch16globecom} \drmub{aimed} is reducing link failure detection delay. For this purpose, authors used bidirectional forwarding detection (BFD) and SDN controller's heart beat (HB) mechanisms. BFD is done at local level while HB is done at the global centralized level. To address link failure recovery, authors suggested to use fast fail over groups (FFG) provided by the OpenFlow protocol. Moreover, authors proposed to use SDN's controller driven recovery in which SDN controller is responsible for the calculation and recovery of alternative paths. Finally, post optimization recovery has been done by identifying the traffic flows which were affected by the link failure. Then, these identified traffic flows were ordered according to priority and then reprocessed by the general routing module. At the final stage, traffic flows having the top most priority are assigned new optimized paths.

To deal with fault tolerance mechanisms i.e., link failure detection, link failure recovery, and post optimization recovery, authors~\cite{Dorsch16globecom} proposed three approaches. The first approach is centralized approach having BFD and FFG. The second approach is decentralized approach having HB with controller. The third approach is the hybrid one. For the evaluation purpose, two scenarios were considered. The first scenario is substation environment, and the second scenario is wide area monitoring, protection, and control. After extensive analysis on the testbed, authors concluded that the centralized approach performs better in recovery time and optimization. Compared to this centralized approach, the controller based HB approach provide better path as the SDN controller has global view of the whole network but this decentralized approach has less recovery time. In comparison with the centralized and decentralized approaches, the hybrid approach outperformed in terms of recovery time and better path but at the cost of more overhead. All these results were evaluated using OpenFlow's FloodLight SDN controller considering IEC 61850 standard specifications. Authors in~\cite{Dorsch18comnet} provided a comprehensive comparison of different fast recovery approaches.

A fault tolerance approach is proposed in~\cite{Rubaye17iot} in which authors evaluated the end-to-end delay and data flow traffic under the presence of a fault. The proposed SDN architecture incurs less end-to-end delay in comparison with the conventional network setting. Authors basically proposed an algorithm which runs on SDN controller for the computation of end-to-end path between SDN controller and SDN switches.

\subsubsection{Link Failure}
In order to guarantee smooth operation of SG, real time communication is required. This real time communication need to be done even with a delay of sub 10 milliseconds. If a communication link gets broken down, first it need to be identified immediately (a.k.a., link failure detection) and then remedial actions such as alternative communication links should be provided (a.k.a., link recovery). All these communication link failure detection and recovery first need to be measured accurately. Hence, the monitoring of timing of these events is important. \mub{Synchronized link interruption and corruption equipment (SLICE) has been proposed by Kurtz et al.~\cite{Kurtz17sgcom}, which is responsible to provide link interruption and synchronization of the SG communication network.} SLICE is a custom build hardware using resistor, and transistor and then connect it with the global positioning system through a general purpose input and output (GPIO) interface. The main purpose of SLICE is to identify interruptions of communication links, thus facilitating link failure detection and recovery. In order to provide alternative link, authors \drmub{suggested} to use SUCCESS SDN controller~\cite{Dorsch14smartgridcom}. \mub{Dorsch et al.~\cite{Dorsch16globecom} also handle the link failure issues in \drmub{SDN-based SGC} and the details can be found in Section~\ref{ftol}.}

\mub{In contrast to SLICE, Gyllstrom et al.~\cite{Gyllstrom14sgc} proposes link failure detection and reporting algorithms for SG network.} A series of algorithms namely APPLESEED has been proposed for PMU network to address with link failure issues. More discussion on these algorithms are provided in Section~\ref{mcapmu}.

\onecolumn
\begin{landscape}
\begin {figure*}
\scriptsize
\centering
\begin{tikzpicture}[
  level 1/.style={sibling distance=40mm},
  level 2/.append style={sibling distance=40mm},
  edge from parent/.style={->,draw},
  >=latex]

\node[root,minimum height=3em,text width=6cm,fill=red!40] {Taxonomy of \drmub{SDN-based SGC}}
  child {node[level 2,xshift=-130pt,text width=3cm,fill=blue!10] (ch1) {\mhr{Advantages of SDN-based SGC}}
  child {node[level 2,yshift=-30pt,text width=3cm] (c1) {SG Resilience \\Sec. \ref{sgres} \\----------------\\\mub{{\tiny \bf Specific SG Problem Addressed} {\tiny \\----------------\\How to react with sudden failures? \\How to react with malicious attacks? \\What if communication link failure occur? \\What if PMU is compromised? \\What if fast failover occur? \\What if fault occur in SG component? \\How to quickly detect the fault? \\How to reduce failure recovery time?\\What if SDN controller fails?}}}}
  child {node[level 2,yshift=-30pt,xshift=-15pt,text width=3cm] (c2) {SG Stability \\ Sec. \ref{sgstab} \\----------------\\\mub{{\tiny \bf Specific SG Problem Addressed} {\tiny \\----------------\\How to stabilize the voltage? \\How to handle the overload? \\How to perform load balancing? \\How to address power defeciency? \\How to enable NMGs communicate?\\What if fluctuations in wind power occur?}}}}
  child {node[level 2,yshift=-30pt,xshift=-30pt,text width=3cm] (c3) {SG Traffic \mub{Optimization} \\ Sec. \ref{sttraf} \\----------------\\\mub{{\tiny \bf Specific SG Problem Addressed}{\tiny \\----------------\\How to forward packets? \\How to ensure fairness in SMs? \\How to aggregate flows?\\How to allocate bandwidth?}}}}
  child {node[level 2,yshift=-30pt,xshift=-45pt,text width=3cm] (c4) {\mub{Other Advantages} \\ Sec. \ref{miscel} \\----------------\\\mub{{\tiny \bf Specific SG Problem Addressed}{\tiny \\----------------\\How to place concentrators in AMI? \\How to handle aggregator point problem? \\Where to place controller?\\How to capture EVs battery status?}}}}}
      child {node[level 2,yshift=-30pt,xshift=-10pt,text width=3cm,fill=blue!10] (c5) {\mub{Architectures of SDN-based SGC} \\ Sec. \ref{arc} \\----------------\\\mub{{\tiny \bf Specific SG Problem Addressed}{\tiny \\----------------\\How to use optical networks, WSNs, vehicles, and cellular network to support SGC?}}}}  
      child {node[level 2,yshift=-30pt,xshift=-10pt,text width=3cm,fill=blue!10] (c6) {\mhr{Routing} \\ \mhr{Sec. \ref{rout}} \\----------------\\\mhr{{\tiny \bf Specific SG Problem Addressed}{\tiny \\----------------\\ \mhr{How message is delivered to: (a) a single destination?} \\ \mhr{(b) a group of nodes?} \\ \mhr{(c) all the nodes?}}}}}  
        child {node[level 2,yshift=-30pt,xshift=-10pt,text width=3cm,fill=blue!10] (c7) {\mub{Security and Privacy in SDN-based SGC} \\ Sec. \ref{secur} \\----------------\\\mub{{\tiny \bf Specific SG Problem Addressed}{\tiny \\----------------\\How to secure AMI network? \\How to secure PMU network? \\How to protect SG from attacks?}}}};

\begin{scope}[every node/.style={level 3}]

\node [below of = c1, xshift=15pt,yshift=-50pt,xshift=-20pt] (c11) {SDN Controller Failure\\\cite{Ghosh16cypss} \mub{(2016)}};
\node [below of = c11,yshift=-20pt] (c12) {Self-healing and Fast Failover \\ \cite{Lin18tsg} \mub{(2018)},\\ \cite{Kurtz16netsoft} \mub{(2016)},\\\cite{Sydney14comnet} \mub{(2014)}};
\node [below of = c12,yshift=-20pt] (c13) {Fault Tolerance\\\cite{Dorsch16globecom} \mub{(2016)}, \\ \cite{Rubaye17iot} \mub{(2017)},\\ \cite{Zhqng16icsn} \mub{(2016)}};
\node [below of = c13,yshift=-25pt] (c14) {Link Failure\\\cite{Dorsch16globecom} \mub{(2016)},\\ \cite{Gyllstrom14sgc} \mub{(2014)},\\ \cite{Dorsch14smartgridcom} \mub{(2014)},\\ \cite{Kurtz17sgcom} \mub{(2017)}, \\\mub{~\cite{Rehmani18etsn}} \mub{(2018)}};
\node [below of = c14,yshift=-40pt] (c15) {Fast Failure Detection, Diagnosis, and Reduction in Recovery Time\\\cite{Dorsch16globecom} \mub{(2016)}, \\ \cite{Zhqng16icsn} \mub{(2016)},\\ \cite{Lopes17im} \mub{(2017)},\\ \cite{Dorsch18comnet} \drmub{(2018)}};
\node [below of = c15,yshift=-35pt,xshift=30pt,text width=4.5cm] (c16) {SG Resilience \\~\cite{Lin18tsg} \mub{(2018)},\\ \cite{Jin17tsg,Maziku17icnc,Rubaye17iot,Ren17tsg} \mub{(2017)}, \\ \cite{Zhqng16icsn,Aydeger16icc,Ghosh16cypss,Genge16ifip} \mub{(2016)}, \\ \cite{Aydeger15nfvsdn,Dong15cpss} \mub{(2015)}};

\node [below of = c2, xshift=15pt,yshift=-60pt,xshift=-15pt] (c21) {Voltage stability guarantee\\ \cite{Dorsch16smartgridcom} \mub{(2016)} };
\node [below of = c21,yshift=-15pt]	 (c22) {Overload Handling\\ \cite{Dorsch16globecom} \mub{(2016)} };
\node [below of = c22,yshift=-25pt] (c23) {Load Balancing and Management\\\cite{Sydney14comnet} \mub{(2014)},\cite{Hannon16cpads} \mub{(2016)}, \\\cite{Nafi16icspcs} \mub{(2016)}, \mub{~\cite{Montazerolghaem18IoT}} \mub{(2018)}};
\node [below of = c23,yshift=-40pt] (c24) {Power Deficiency and Its Recovery is Achieved through SDN by connecting NMGs\\\cite{Ren17ae} \mub{(2017)}};
\node [below of = c24,yshift=-30pt] (c25) {Stability against fluctuations in wind power\\\cite{Rayati18tii} \mub{(2018)}};

\node [below of = c3, xshift=15pt,yshift=-70pt,xshift=-20pt] (c31) {SG Traffic Recovery and Optimization\\ \cite{Dorsch16globecom} \mub{(2016)} };
\node [below of = c31,yshift=-30pt]	 (c32) {Scheduling and Flow Aggregation\\ \cite{Guo16wimob} \mub{(2016)}, \\ \cite{Singh17jpdc} \mub{(2017)}, \\ \cite{Molina15caee} \mub{(2015)},\cite{Ngyuen13ic} \mub{(2013)} };
\node [below of = c32,yshift=-25pt] (c33) {Fairness Among Smart Meters\\\cite{Guo16wimob} \mub{(2016)}};
\node [below of = c33,yshift=-15pt] (c34) {Packet Forwarding Performance\\\cite{Kurtz16netsoft} \mub{(2016)}};
\node [below of = c34,yshift=-25pt] (c35) {Throughput, QoS, and BW Allocation\\\cite{Guo16wimob} \mub{(2016)}, \cite{Sydney14comnet} \mub{(2014)}, \\ \cite{Li16sege} \mub{(2016)}};
\node [below of = c35,yshift=-25pt] (c36) {Substation traffic communication\\\cite{Dorsch14infocom} \mub{(2014)}};

\node [below of = c4, xshift=15pt,yshift=-70pt,xshift=-20pt] (c41) {Concentrator and Controller Placement Problem\\ \cite{Guo16iccc} \mub{(2016)}, \\ \cite{Wang16infocom} \drmub{(2016),} \\ \cite{Nafi17fgcs} \mub{(2017)} };
\node [below of = c41,yshift=-25pt]	 (c42) {Network \\ Expansion\\ \cite{Zhqng16icsn} \mub{(2016)} };
\node [below of = c42,yshift=-15pt] (c43) {CPU Utilization\\\cite{Zhqng16icsn} \mub{(2016)}};
\node [below of = c43,yshift=-15pt] (c44) {Battery Status Sensing\\\cite{Li17sensors} \mub{(2017)}};
\node [below of = c44,yshift=-15pt] (c45) {State Estimation of Electric Grid\\\cite{Meloni17energies} \mub{(2017)}};
\node [below of = c45,yshift=-15pt] (c46) {Automatic Network Reconfiguation\\\cite{Deutsch18GioTs} \drmub{(2018)}};
\node [below of = c46,yshift=-15pt] (c47) {Monitoring of SDN in SG\\\cite{Rinaldi15wfcs} \drmub{(2015)}};

\node [below of = c5, xshift=20pt,yshift=-70pt,xshift=-10pt,text width=3cm] (c51) {Applied to Optical Networks \\ Sec. \ref{sdnopt}\\-------------------------\\Hybrid Opto-Electronic Ethernet \footnotesize \cite{Zheng16smartgridcom} \mub{(2016)}\\-------------------------\\Optical Transmission Reliability \footnotesize \cite{Rastegarfar16icnc} \mub{(2016)}};
\node [below of = c51,yshift=-60pt,text width=3cm]	 (c52) {Applied to WSNs \\ Sec. \ref{sdwsnar}\\-------------------------\\SG-NAN  \\  \footnotesize \cite{Nafi16icspcs} \mub{(2016)}\\-------------------------\\Sensor Open Flow  \\  \footnotesize \cite{Sayyed14ice} \mub{(2014)} };
\node [below of = c52,yshift=-70pt,text width=3cm] (c53) {Applied to Vehicles \\ Sec. \ref{sdnvehar}\\-------------------------\\V2G  \\  \footnotesize \cite{Zhang16sege} \mub{(2016)}, \\ \cite{Li17sensors,Chen17network} \mub{(2017)}, \\ \cite{Sun17commag} \mub{(2017)}, \\ \cite{Chekired18tii,Hu18Commag,Wang18Itsm} \mub{(2018)}\\-------------------------\\ G2V  \\  \footnotesize \cite{Nafi16icspcs} \mub{(2016)}};
\node [below of = c53,yshift=-60pt,text width=3cm] (c54) {Applied to Cellluar Networks \\ Sec. \ref{sdnlte}\\-------------------------\\LTE  \\ \footnotesize \cite{Rubaye17wcom} \mub{(2017)}};


\node [below of = c6, yshift=-90pt,xshift=5pt,text width=3cm] (c61) {Unicast \\ Sec. \ref{unrtr} \\-------------------------\\ \footnotesize \cite{Germano15im}   \mhr{(2015)}, \\ \footnotesize \cite{Zhqng16icsn}  \mhr{(2016)}, \\ \footnotesize \cite{Zhao2016}  \mhr{(2016)}, \\ \footnotesize \cite{Alharbi16smartgridcom}   \mhr{)2016)}, \\ \footnotesize \cite{Patil16nca}   \mhr{(2016)}, \\ \footnotesize \cite{Genge16ifip}  \mhr{(2016)}, \\ \footnotesize \cite{Han17Globecom}   \mhr{(2017)}, \\ \footnotesize \cite{Rubaye17iot}   \mhr{(2017)}, \\ \footnotesize \cite{Kaur18Tkde}   \mhr{(2018)} };
\node [below of = c61,yshift=-105pt,text width=3cm]	 (c62) {Multicast \\ Sec. \ref{multiSDN-based SG} \\-------------------------\\ \footnotesize \cite{Li17sensors}  \mhr{(2017)}, \\ \footnotesize \cite{Gyllstrom14sgc}  \mhr{(2014)}, \\ \footnotesize \cite{Goodney13smartgridcom}   \mhr{(2013)}, \\ \footnotesize \cite{Montazerolghaem18IoT}   \mhr{(2018)}, \\ \footnotesize \cite{Zheng16smartgridcom}  \mhr{(2016)}, \\ \footnotesize \cite{Dorsch14smartgridcom}   \mhr{(2014)}, \\ \footnotesize \cite{Pfeiff15ntms}  \mhr{(2015)}, \\ \footnotesize \cite{Lopes15ac}  \mhr{(2015)}, \\ \footnotesize \cite{Zhou18Tii}  \mhr{(2018)} };
\node [below of = c62,yshift=-95pt,text width=3cm] (c63) {Multicast/Broacast \\ Sec. \ref{rtmb} \\-------------------------\\ \footnotesize \cite{Alishahi2016p2p}   \mhr{(2016)}, \\ \footnotesize \cite{Kim15smartgridcom}   \mhr{(2015)}   };


\node [below of = c7, yshift=-90pt,xshift=5pt,text width=3cm] (c71) {Applied to SubStation \\ Sec. \ref{subst}\\-------------------------\\IDS \\ \footnotesize \cite{Ghosh17icdcsw} \mub{(2017)},\\ \cite{Silva16compsac} \mub{(2016)}\\-------------------------\\Link Flood Attack \\ \footnotesize \cite{Maziku17icnc} \mub{(2017)}\\-------------------------\\Anti-Eavesdropping \\
        \footnotesize ~\cite{Germano15im} \mub{(2015)}\\-------------------------Authentication Scheme \\
        \footnotesize \mub{~\cite{Aydeger18ccnc}} \mub{(2018)}};
\node [below of = c71,yshift=-105pt,text width=3cm]	 (c72) {Applied to AMI and PMU Networks \\ Sec. \ref{amipm}\\-------------------------\\AMI  \\ \footnotesize \cite{Irfan15cit} \mub{(2015)}, \cite{Whang17mis} \mub{(2017)}, \mub{~\cite{Chi14Secureware}} \mub{(2014)}, \mub{~\cite{Aujla18Tii}} \mub{(2018)}\\-------------------------\\PMU  \\ \footnotesize \cite{Jin17tsg} \mub{(2017)}, \cite{Lin18tsg} \mub{(2018)}};
\node [below of = c72,yshift=-95pt,text width=3cm] (c73) {Applied to Different Networks \\ Sec. \ref{diffnet}\\-------------------------\\V2G  \\ \footnotesize \cite{Zhang16sege} \mub{(2016)}\\-------------------------\\VLANs  \\ \footnotesize \cite{Kim14smartgridcom} \mub{(2014)}\\-------------------------\\\mub{General SG Systems}  \\ \footnotesize \mub{~\cite{Ibdah17icecta}} \mub{(2017)}, \\ \mub{~\cite{Danzi18arxiv}} \mub{(2018)} };

\end{scope}

\foreach \value in {1,2,3,4,5,6}
  \draw[->] (c1.195) |- (c1\value.west);

\foreach \value in {1,2,3,4,5}
  \draw[->] (c2.198) |- (c2\value.west);

\foreach \value in {1,2,3,4,5,6}
  \draw[->] (c3.198) |- (c3\value.west);

\foreach \value in {1,2,3,4,5,6,7}
  \draw[->] (c4.198) |- (c4\value.west);

\foreach \value in {1,2,3,4}
  \draw[->] (c5.198) |- (c5\value.west);
  
  \foreach \value in {1,2,3}
  \draw[->] (c6.198) |- (c6\value.west);  
  
  \foreach \value in {1,2,3}
  \draw[->] (c7.198) |- (c7\value.west);

\end{tikzpicture}
\caption {Taxonomy of SDN-based \drmub{SGC}. Advantages of SDN-based SGC are classified into the left four branches (cf. Sec. \ref{sgres}, Sec. \ref{sgstab}, Sec. \ref{sttraf}, and Sec. \ref{miscel}) according to four areas namely resilience, stability, traffic \mub{optimization}, and under \mub{other advantages} category. We have classified SDN-based SGC architectures, routing, and security and privacy schemes as well, see right two branches (cf. Sec. \ref{arc}, Sec. \ref{rout}, and Sec. \ref{secur}).}
\label{fadvsdnsg}
\end{figure*}
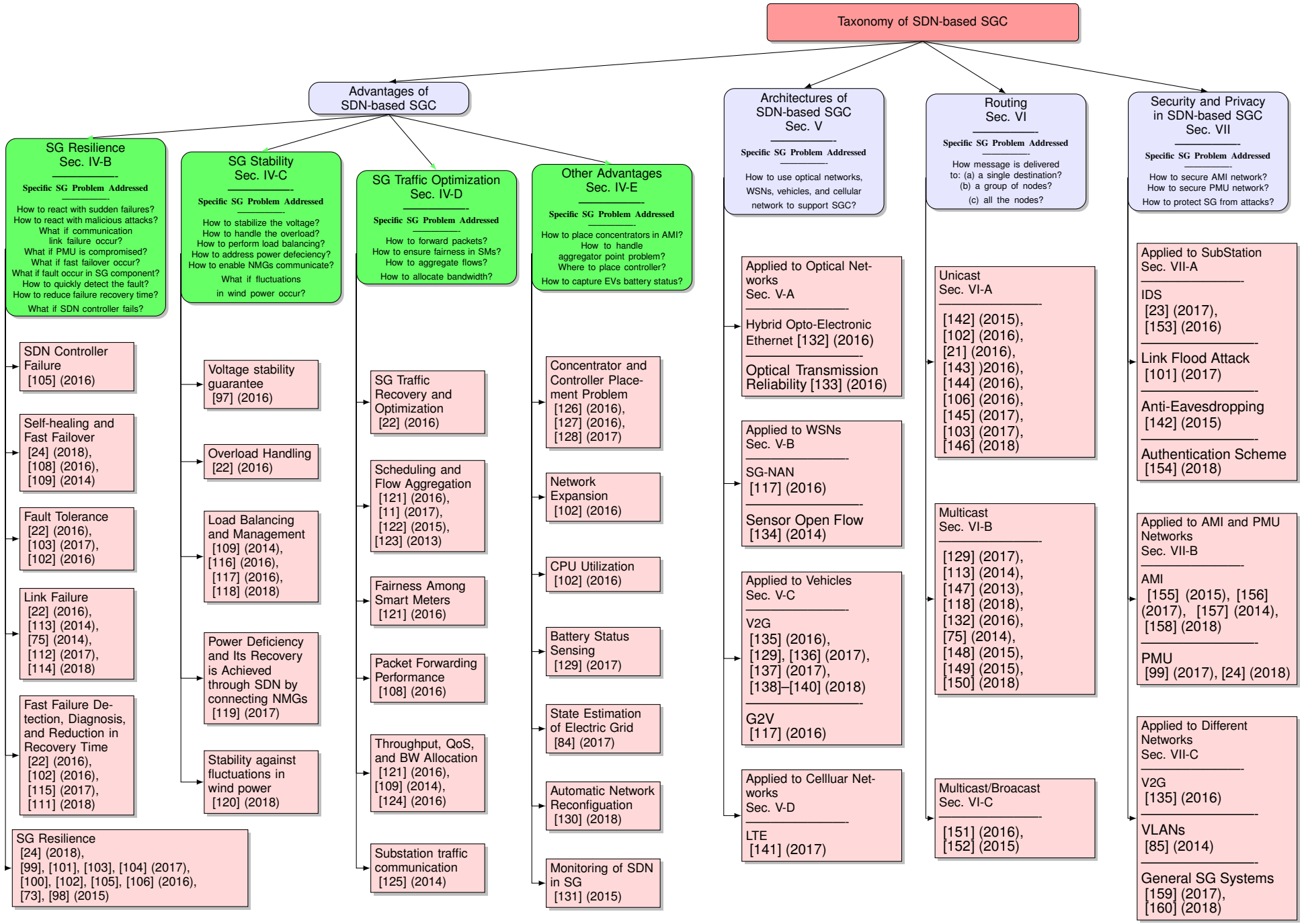

\end{landscape}
\twocolumn

\begin{figure*}[htbp]
	\centering
	\begin{subfigure}[]{}
		\centering
		\includegraphics[width=0.45\textwidth]{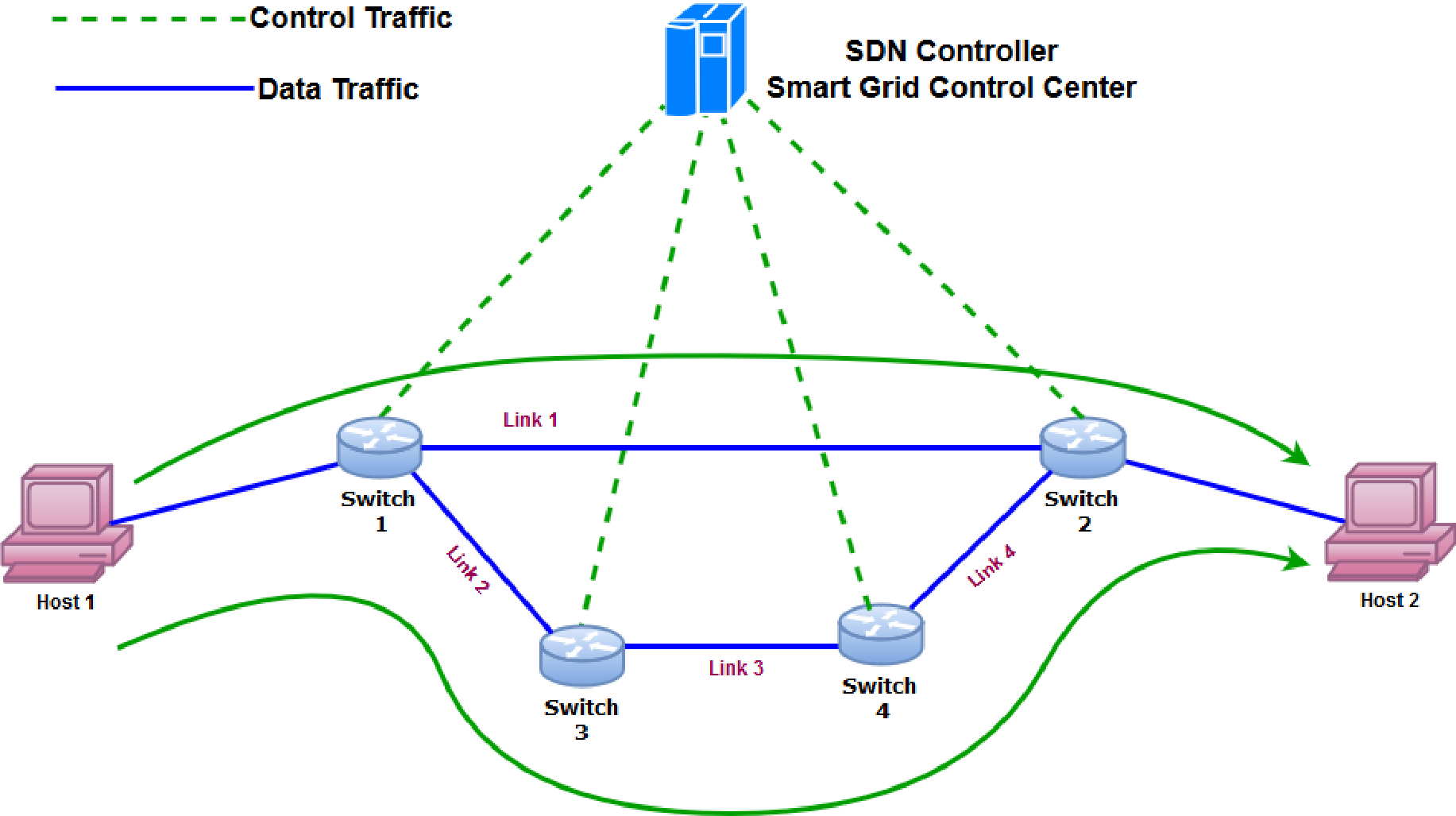}
	\end{subfigure}%
	\qquad
	\begin{subfigure}[]{}
		\centering
		\includegraphics[width=0.45\textwidth]{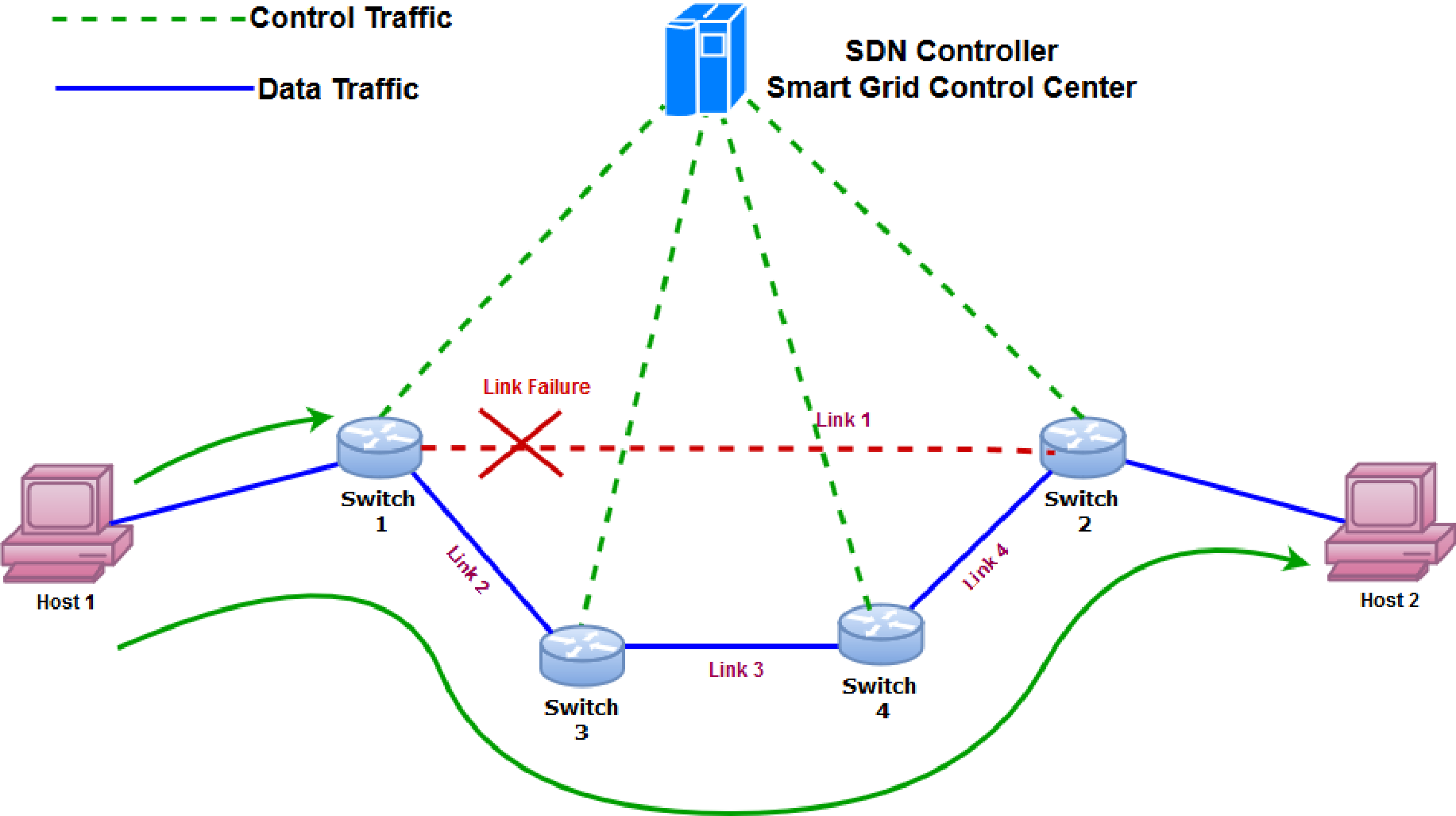}
	\end{subfigure}%
	\caption{\mub{Illustration of a link failure scenario in a simple network topology showing SDN controller present in the \drmub{SG} control centre having four switches. There are two communication paths which may be adopted for communication when: (a) without link failure, and (b) with link failure. The LFL-MAB algorithm helps the SDN controller to learn link failures through advanced machine learning technique\protect\cite{Rehmani18etsn}.}} 
	\label{fig:lflmab}
\end{figure*}

\mub{In SG environment, it may possible that an attacker may compromise SG devices and then launch link failure attacks, as shown in Fig.~\ref{fig:lflmab}. To handle this situation, a link failure learning algorithm using multi-armed bandit theory (LFL-MAB) algorithm has been proposed in~\cite{Rehmani18etsn}. LFL-MAB algorithm has been tested under various attacking modes and validated through Mininet based simulations considering RYU controller and OpenFlow switches. Compared to centralized and random approach, LFL-MAB strategy is able to learn the link failure attacking strategy selected by an SG attacker and is able to switch to more reliable communication links.}

\subsubsection{Multiple Failure Scenario}
In an \drmub{SG} environment, multiple failure scenarios can occur simultaneoulsy. This will severly deteriorate the performance of the SG communication system. \mub{Dorsch et al.~\cite{Dorsch14smartgridcom} discusses these multiple failure scenarios studies where three cases for manufacturing message specification (MMS) traffic were considered.} We ask the reader to refer Section~\ref{rutsbs} for more details. 

\subsubsection{Quick Fault Detection}
\label{qfd}

\mub{Zhang et al.~\cite{Zhqng16icsn} considers synchronous digital hierarchy (SDH) based SG network in which SDN technology is used for quick fault detection, and network expansion scenario in the SG.} Authors considered SDN based NOX controller which manages the switches in the whole network. End-to-end path provisioning in SDH based SDN network is achieved under normal and network expansion scenarios. It is demonstrated that the proposed approach quickly diagnose the fault as well as provide capability to achieve end-to-end path restoration. Similarly, another  study~\cite{Dorsch16globecom} also handle the link failure issues in \drmub{SDN-based SGC} and the details can be found in Section~\ref{ftol}.

\subsubsection{Reduction in Failure Recovery Time}
A SDN based framework, ARES, is proposed in~\cite{Lopes17im} to reduce the failure recovery time in SG. ARES framework is designed in such a way that it meets the SG protection requirements from generation to distribution. In fact, a new API is introduced in ARES which allows the users to modify the network dynamically. This new layer can help introduce new control services in SG at the management layer i.e., SCADA-NG. This SCADA-NG is very flexible and robust. Moreover, SCADA-NG can have the information of the whole SG network, thus it can easily perform failure recovery. Furthermore, SCADA-NG will also keep track of all the incoming and outgoing devices. Authors evaluated the ARES framework by using Mininet, OpenFlow and RYU controller. The comparison of ARES is done with rapid spanning tree protocol (RSTP) and it is shown that the failure recovery time is reduced to micro seconds which is the essential requirement for SG protection.

\subsubsection{SDN Controller Failure}
\label{sdnctrlfail}

The research on applying SDN for SG communication has gained significant momentum. As discussed in Section~\ref{sdn}, the SDN controller will have a global network view and instructs the switches to perform specific tasks. This SDN controller is generally a single entity in the network which may become a single point of failure. SDN controller faults may belong to different categories. For instance, an SDN controller may stop working due to various hardware or software irregularities. It may happen that a SDN controller may become resource constraint due to several request by a malicious switch, causing denial of service by the SDN controller. It may also happen that communication link failure occurs between the SDN controller and the switches. Furthermore, a malicious user may inject a malware in the SDN controller itself, thus causing the SDN controller to misbehave. Finally, due to any software problem, any application running on SDN controller may become faulty and instruct the switches in a bad manner. There may be other types of faults faced by the SDN controller such as a SDN controller may get hang or a SDN controller may modify or delete the entry in the flow tables of the switches.

All these SDN controller faults will either results in random failure of the SDN controller or delayed response by the SDN controller. If the SDN controller fails completely, it will severely affect the SG communication network. Even if the SDN controller is not completely failed, and it may respond with a non-negligible delay, it may cause severe packet drops by the switches. In fact, every packet in a switch need to see the flow entry within a time limit and then take the corresponding action. If the SDN controller does not provide a timely flow entry to instruct a particular flow or type of packets, these packets may get directed to the wrong flow entry or the packets may get dropped. The dropped packets may have very critical impact on the SG as they may contain control traffic or commands given by the SDN controller. Thus, the time critical communication in SG may suffer~\cite{Ghosh16cypss}.

\mub{In an effort to deal SDN controller failures and its impact on SG, Ghosh et al.~\cite{Ghosh16cypss} presented a detailed study.} More precisely, the authors focused on  the impact of SDN controller failure of automatic gain control (AGC) of SG communication system. AGC maintains and regulates the frequency of the grid. IEEE 37 bus system along with Mininet and PowerWorld simulation tools were used to study the impact of SDN controller failures. Authors also used real SDN switches to study this impact. Through extensive simulations and hardware measurement results, authors demonstrated that the SDN controller failure severely degrade the AGC performance in SG.

\subsection{SG Stability}
\label{sgstab}

\subsubsection{Voltage Stability}
\mub{For voltage stability, Dorsch et al.~\cite{Dorsch16smartgridcom} proposes to use multi agent system (MAS) in conjunction with SDN in substation environment.} The agent is deployed at each substation of the \drmub{SG}. SDN NBI is implemented through which control agents communicate with the SDN controller directly. The SDN controller makes forwarding rules and establishes them whenever control agents communicate them to the SDN controller. In this manner, power grid voltage stability is achieved by timely and reliably transmitting critical control messages and commands in a substation environment. The proposed MAS SDN system is evaluated considering IEC 61850 standard. The SDN controller worked as REST server. The purpose of REST server is to provide services through dedicated URLs. Moreover, to interact between MAS and SDN controller, SDN controller NBI is implemented using RESTful API. Additionally, Java based Floodlight controller using OpenFlow is considered. 

\subsubsection{Overload Handling}
In a \drmub{SG} environment, it may happen that some links are more overloaded than the others. This will cause congestion issues over the link and the ultimate result will be in the shape of packet loss. \mub{To address this issue, Dorsch et al.~\cite{Dorsch16globecom} uses SDN approach for this type of link overload.} For instance, in a post recovery optimization scenario, through SDN controller, less overloaded links will be selected. 

\subsubsection{Load Balancing and Management}

\mub{Hannon et al.~\cite{Hannon16cpads} manages load balancing by proposing a load shifting algorithm and managed the load of the \drmub{SG} through SDN efficiently.} In fact, authors proposed distribution system solver network (DSSnet) testbed, which basically plays the role of simulator as well as emulator. The unique feature of DSSnet is that it not only models the typical power system but it also enables to test the impact of IEDs over the system. Moreover, with DSSnet, one can interact with the power system and communication network system simultaneously. Furthermore, DSSnet by incorporating SDN features such as programmable switches and SDN controller, along with Mininet integration, helps the researchers to analyse their proposed algorithms effectively. DSSnet system integrates two systems together, one is the power system, which runs on Windows based machine, \mub{and the other one is Mininet emulator,} which runs on Linux environment. DSSnet has been built to test and validate SDN based communication network, flow of power in the system, and several SG applications built on top of it. Synchronization among the events and between the two systems is achieved by using the concept of virtual time. 

\mub{In the context of load balancing in AMI network using SDN controller, OpenAMI routing scheme is proposed in~\cite{Montazerolghaem18IoT}. OpenAMI achieves low end-to-end delay and higher throughput by selecting shortest route and balancing the traffic load in the entire AMI network (cf. Section~\ref{raminet} for more details of this work).}

\subsubsection{Power Deficiency and Its Recovery}
Microgrids are getting more attention both from the research community and utilities due to their obvious advantages. These advantages include:  (a) easier integration and management of \drmub{RERs}, (b) microgrids facilitate customers in a satisfactory manner, and (c) microgrids are cheaper in terms of economy and cause less emissions~\cite{Ren17ae}. A microgrid (as also discussed in Section~\ref{pmunet}) can be connected to the main utility power grid a.k.a., grid connected mode or it can work on standalone basis a.k.a., islanded mode. A commercial building or a small residential area or even university campus can be considered as a microgrid. With these aforementioned advantages, microgrids are getting deployed at a faster pace in urban areas. In future smart cities~\cite{Mouftah18book}, microgrids will be the essential requirement. However, the main issue is to coordinate and managed these small scale networked microgrids (NMGs). For instance, it may happen that at a particular instance of time, one microgrid has excess of energy, while the other neighboring microgrid needs some energy. With effective management and coordination, energy can be easily transferred and shared among these NMGs but a reliable communication system is required. \mub{To address this communication problem of NMGs, Ren et al.~\cite{Ren17ae} proposes to use SDN technology.} More precisely, power deficiency and its recovery is achieved through SDN by connecting the closely coupled NMGs. The SDN controller will decide which links to establish for communication between NMGs based upon a triggered event. Authors used OP5600 OPAL-RT simulator and RYU is used as SDN controller.

\subsubsection{Stability Against Fluctuations in Wind Power}
Integrating wind power with the \drmub{SG} brings instability in the power grid to some extent. This is primarily because wind power generation is not constant always and impacts the frequency of the power grid system. \mub{In order to deal with this situation, Rayati et al.~\cite{Rayati18tii} proposes stability achievement against fluctuations in wind power by proposing a control system.} The proposed control system is merged with the cloud and SDN technology is used for timely communication and to preserve privacy over the cloud. 

\subsection{\mub{SG Traffic Optimization}}
\label{sttraf}

SDN can bring lot of advantages in the management of SG traffic \mub{optimization}. For instance, SDN's controller can help to build different data trees depending upon specific SG application and its QoS requirements. These data trees can work in publish/subscribe fashion. The utility needs data from PMU and meters residing in consumer's premises. With the help of SDN, two different data trees with varying depth and QoS requirements can be easily managed. The PMU's data collection tree can be of real-time QoS requirement with smaller depth, while consumer's meter data collection tree can be of smaller width in order to tackle with limited memory constraint of flow tables in the OpenFlow switches. Below we discuss all these traffic related advantages of \drmub{SDN-based SGC}.

\subsubsection{Packet Forwarding Performance}
Communication requirements for critical infrastructure such as \drmub{SG} will also be the part of 5G (like support of machine type communication). In this context, an architectural concept for \drmub{SG} communication for 5G has been presented in~\cite{Kurtz16netsoft}. For the use case, authors developed a testbed. This testbed, ``SDN4CriticialInfrastructure'', consists of eleven Ethernet switches, one FloodLight OpenFlow controller, and six PCs for traffic generation. Authors created a topology in the testbed to measure packet forwarding performance in terms of delay. Both the performance of forwarding latencies for Bare-Metal switch and virtual switch were evaluated by varying the number of packets from 500 packets/second to 8000 packets/second as provided by the IEC 61850 standard. It is shown that Bare-Metal switch outperformed virtual switch in terms of latencies irrespective of packet sending rate. It is also shown that virtual switches are far better in restoring the communication links if any link failure occur. We ask the readers to refer~\cite{Kurtz16netsoft} for a detail comparison of Bare-Metal switches and virtual switches.

\subsubsection{Fairness Among Smart Meters}
\mub{Guo et al.~\cite{Guo16wimob} exploits SDN's flow-level management feature and perform aggregation and scheduling of traffic flows to achieve \mub{fairness} in smart meters.} In fact, an SDN-based framework is proposed for the smart meters so that their throughout increases. More precisely, the goal is that each smart meter get a uniform share at the flow level. Authors used NS-3 and Mininet based evaluation and it is demonstrated through extensive results that fairness is achieved in smart meters. Jain's index is used for measuring the fairness index.

\subsubsection{Scheduling and Flow Aggregation}
\mub{Guo et al.~\cite{Guo16wimob} considers the scheduling and flow aggregation for smart meters using SDN. In an effort to use SDN technology for IEC 61850 substation and perform flow management, OpenFlow protocol, and Floodlight controller was used by Molina et al.~\cite{Molina15caee}.} Authors evaluated the proposed flow aggregation scheme through Mininet and showed that load balancing, scheduling, and flow aggregation can be easily achieved through SDN.

\subsubsection{Bandwidth Allocation}
\mub{Li et al.~\cite{Li16sege} proposes a bandwidth allocation scheme for devices which operates in IEC 61850 substation setting.} For the evaluation purpose, authors used OpenFlow, Mininet, and OpenvSwitch and demonstrated that with the help of SDN controller, the bandwidth improvement can be till 90\%. 

\subsubsection{WAN Traffic Communication Applications}
Real time communication capabilities of SG communication network is analyzed for three substation scenarios in~\cite{Dorsch14infocom}. In the first scenario, communication within the substation is analyzed. In the second scenario, communication between substations is analyzed. And finally in the third scenario, communication between substation and utility control station is analyzed. In all the aforementioned communication scenarios, IEC 61850 standard was considered. IEC 61850 supports three major communication services named as sampled values (SV), generic object oriented substation event (GOOSE), and manufacturing message specification (MMS). 

The IEC 61850 standard was initially introduced for substation communication in 2006 by International Electrotechnical Commisssion (IEC) technical committee 57~\cite{iecstd}. IEC 61850 is widely accepted in the industry as well and it has been used even in the oil and gas industry for the control and monitoring of their electrical systems~\cite{Montignies11iam}. The goal of IEC 61850 is to facilitate communication within substation using interoperable intelligent electronic devices (IEDs). Later in 2009~\cite{iecttc57} and 2013~\cite{iectc57control}, IEC 61850 standard is upgraded for inter-substation communication and substation to utility control center communications. Moreover, IEC 61850 has also the capabilities to incorporate EVs and DERs support. 

\subsection{\mub{Other Advantages}}
\label{miscel}

\subsubsection{\drmub{Concentrator and Controller Placement Problem: }}

\drmub{AMI} network plays a vital role in future SG. In AMI, smart meters send information to the utility through concentrators. Similarly, the utilities communicate real time prizing information and firmware updates to the smart meters (customers) through the concentrators. These concentrators in AMI network are basically the relay point. If a large number of smart meters are associated with a single concentrator, there may be congestion issues and if less number of smart meters are assigned to the concentor, there will be under utilization of the resources. Thus, plancement of concentrator within an AMI network is a serious issue. \mub{Guo et al.~\cite{Guo16iccc} addresses this issue by proposing a genetic algorithm for the optimal placement of concentrator in AMI network.} However, it is not clear that where SDN has been used except the fact that the SDN terminology was used in the title.

In AMI network, aggregation points will be deoplyed to handle the data between the smart meters and utility. \mub{Wang et al.~\cite{Wang16infocom} handles the aggregation point's problem and try to minimize the number of aggregation points to be deployed in urbran, sub-urban, and rural scenarios.} Though authors suggested to use \drmub{South Bound Interface (SBI)} for control plane and data plane communication of SDN, but other exact features of SDN were heavily missing in the paper.

\mub{Nafi et al.~\cite{Nafi17fgcs} focuses on controller placement problem in a SDN based WSN for NAN.} Authors considered different SG applications (distribution automation, \drmub{DR}, reading from smart meters, and outage management) and evaluated their proposed algorithms through Castalia, which is a simulator based on OMneT++ platform.  

\subsubsection{\drmub{CPU Utilization/Network Expansion: }}

The problem of network expansion and its impact on CPU Utilization has been discussed and evaluated in~\cite{Zhqng16icsn}. Authors used SDN technology for SDH based SG networks where fault tolerance is achieved. More details can be found in Section~\ref{rgen} and Section~\ref{qfd}.

\subsubsection{\drmub{Battery Status Sensing: }}

Battery status sensing architecture for software defined vehicle-to-grid (SD-V2G) architecture is proposed in~\cite{Li17sensors}. In fact, a multicast scheme for battery status sensing based on \drmub{SDNs} (BSS-SDN) is proposed which try to reduce the average delay for different V2G services. 

\subsubsection{\drmub{Automatic Network Reconfiguration: }}
\drmub{It normally happens in SG that devices are relocated at different locations, thus trigging node discovery at the new location and subsequent update at the SDN controller will take place. This also results in the re-routing of information to the new associated substation controller. In order to trigger automatic network reconfiguration, Tobias et al.,~\cite{Deutsch18GioTs} proposed to simplify the application level deployment of the SGC application. In fact, Tobias et al. analysed a use case in a substation environment in which the utility (grid operator) changes the feeder configuration as soon as any device (voltage sensors for example) changes it's association and physically connected to a different substation by making changes in the SDN itself.}

\subsubsection{\drmub{State Estimation: }}
State estimation of \drmub{SG} through SDN is presented in~\cite{Meloni17energies}. In a \drmub{SG} wide area monitoring system, PMUs are installed which regularly updates the current state of the grid. In some cases, the resources (PMUs and IEDs) need to scale-up and scale-down, depending upon the monitoring requirements. Thus, SDN can help in this context through which utilities can easily include or exclude the PMUs without any manual intervention in the grid. Authors considered IEEE 14-bus test network for evaluation purpose and used Mininet to validate the proposed bandwidth allocation algorithm for monitoring devices. Moreover, RYU controller along with OpenFlow protocol is used and ofsoftswitch13 were also used as virtual switches.

\subsubsection{\drmub{Monitoring of SDN in SG: }}
\mub{For the monitoring of SDN in SG heterogeneous environment (having IEEE 802.11 link, Gigabit Ethernet link, and Broadband Power Line), Rinaldi et al.~\cite{Rinaldi15wfcs} presents preliminary results of the testbed.} Authors demonstrated that SDN is a good choice for SG monitoring applications and compared software-based SDN switch with hardware-based SDN switch. 

\drmub{MPLS} has been adopted by utilities in the early stages as it provides traffic engineering and the support of creating virtual private networks (VPNs). MPLS is based on routing protocols such as resource reservation protocol (RSVP) or open shortest path first (OSPF). However, relying on MPLS restricts the researchers to implement and test new \mub{protocols} and services on real SG environment. Even improving MPLS requires to manually configure the routers at a massive scale in the SG network, which is not a feasilbe long term solution. \mub{In order to address these issues, Sydney et al.~\cite{Sydney13tsg} performs the simulative comparison between MPLS and OpenFlow protocol and advocates that OpenFlow provides the same features adn performance as of MPLS.} Besides this, OpenFlow enables the utilities and researchers to test their newly created protocols easily. For simulating the power system, authors used toolkit for hybrid systems modeling and evoluation (THYME)~\cite{thyme18}. Network simulator (NS-3)~\cite{ns3} has been adopted to simulate the communication network of SG. Since, both THYME and NS-3 are developed in C++, therefore, their integration was easier. Moreover, NS-3 has both the modules of MPLS routers and OpenFlow switches. During the simulations, a sample SG application was considered in which control commands were issued to actuate the control system for managing the generator speeds. These applications were tested in IEEE 118-bus and IEEE 300-bus topologies. 

\subsection{\mhr{Summary and Lessons Learned}}
\mhr{In this section, we have surveyed taxonomy of SDN-based SGC. First, the motivations of adopting SDN in the SG were discussed. Then we presented advantages of SDN-based SGC in SG resilience, SG stability, and SG traffic optimization. Finally, we also discussed other advantages of SDN-based SGC. We noticed that SDN-based SGC has been widely used in the aforementioned areas of resilience, stability, and traffic optimization, however, there are few other areas where advantages can be seen implicitly. For instance, when SDN-based SGC is used for unicast, multicast, and broadcast message delivery, it outperformed traditional ways of routing. Moreover, there are few areas where the advantages are not fully explored. Therefore, further investigation is required to explore the areas of network expansion, battery status sensing, and state estimation of SG. Moreover, autonomic network configuration is another venue for future research direction which needs to be explored in the context of SG.}

\section{Architectures of \drmub{SDN-based SGC}}
\label{arc}

In this section, we discuss architectures presented so far in the literature for \drmub{SDN-based SGC}. \drmub{The right branch of Fig.~\ref{fadvsdnsg} shows architectures of SDN-based \drmub{SGC}.} These architectures are proposed for optical networks, wireless sensor networks, and \drmub{vehicle-to-grid (V2G)} or \drmub{grid-to-vehicle (G2V)} networks. \mub{Jaraweh et al.~\cite{Jararweh15aiccsa} proposed a general SDN based SG architecture.} In the proposed architecture, the authors \drmub{advocated} to use different software defined systems such as software defined Internet of Things, software defined storage, software defined security, among others and the goal is to provide a secure and reliable SG. However, authors did not validate their proposed architecture and no proof of concept is provided to validate it.

\subsection{\drmub{SDN-based SGC} architectures applied to Optical Networks}
\label{sdnopt}

In this sub-section, we discuss \drmub{SDN-based SGC} architectures applied to optical networks. 

\subsubsection{Hybrid Opto-Electronic Ethernet}
This study~\cite{Zheng16smartgridcom} propose to use hybrid opto-electronic Ethernet network architecture based on SDN for substation communication. Considering the IEC 61850 standard sample value (SV) type 4 messages, the goal is to reduce the delay which is essential for substation communication. We describe more about routing through SDN used in this architecture later in Section~\ref{rton}.

\subsubsection{Optical Transmission Reliability}
\mub{To achieve optical transmission reliability in SG, Rastegarfar et al.~\cite{Rastegarfar16icnc} proposes to used SDN.} Authors basically \drmub{proposed} a SDN based cyber physical system (CPS) interdependency model which uses optical fiber links for the control messaging within the SG. The goal is to avoid cascading failure problems from the CPS perspective through SDN.  The proposed approach is evaluated using 24 nodes US and 28 node EU physical topology and it is shown that number of nodes failures were comparatively less.

\subsection{\drmub{SDN-based SGC} architectures applied to Wireless Sensor Networks}
\label{sdwsnar}

In this sub-section, we discuss \drmub{SDN-based SGC} architectures applied to wireless sensor networks.

\subsubsection{SG-NAN}
A SG-NAN architecture for grid-to-vehcile load management is proposed in~\cite{Nafi16icspcs} using software defined wireless sensor networks (SDWSNs). This architecture is discussed in more detail later in Section~\ref{sdnsgg2v}.

\subsubsection{Sensor Open Flow}
\mub{``Sensor Open Flow (SOF)'', a software defined wireless sensor network architecture for \drmub{SG} applications has been presented by Sayyed et al.~\cite{Sayyed14ice}.} The architecture consists of application layer, control plane, and the data plane. The application layer have customized applications with respect to \drmub{SG}. The control plane will have an SDN controller which will interact with the programmable sensor nodes through SOF. And finally, the data plane contains numerous wireless sensor nodes. Each wireless sensor node will have a flow table contains entries provided by the SOF. Authors suggested to use OpenFlow protocol to evaluate their proposed architecture, however, the paper presents this initial architecture without proof of concept and simulations results.  

\begin{figure*}[t] \centering
\includegraphics[width=0.6\textwidth]{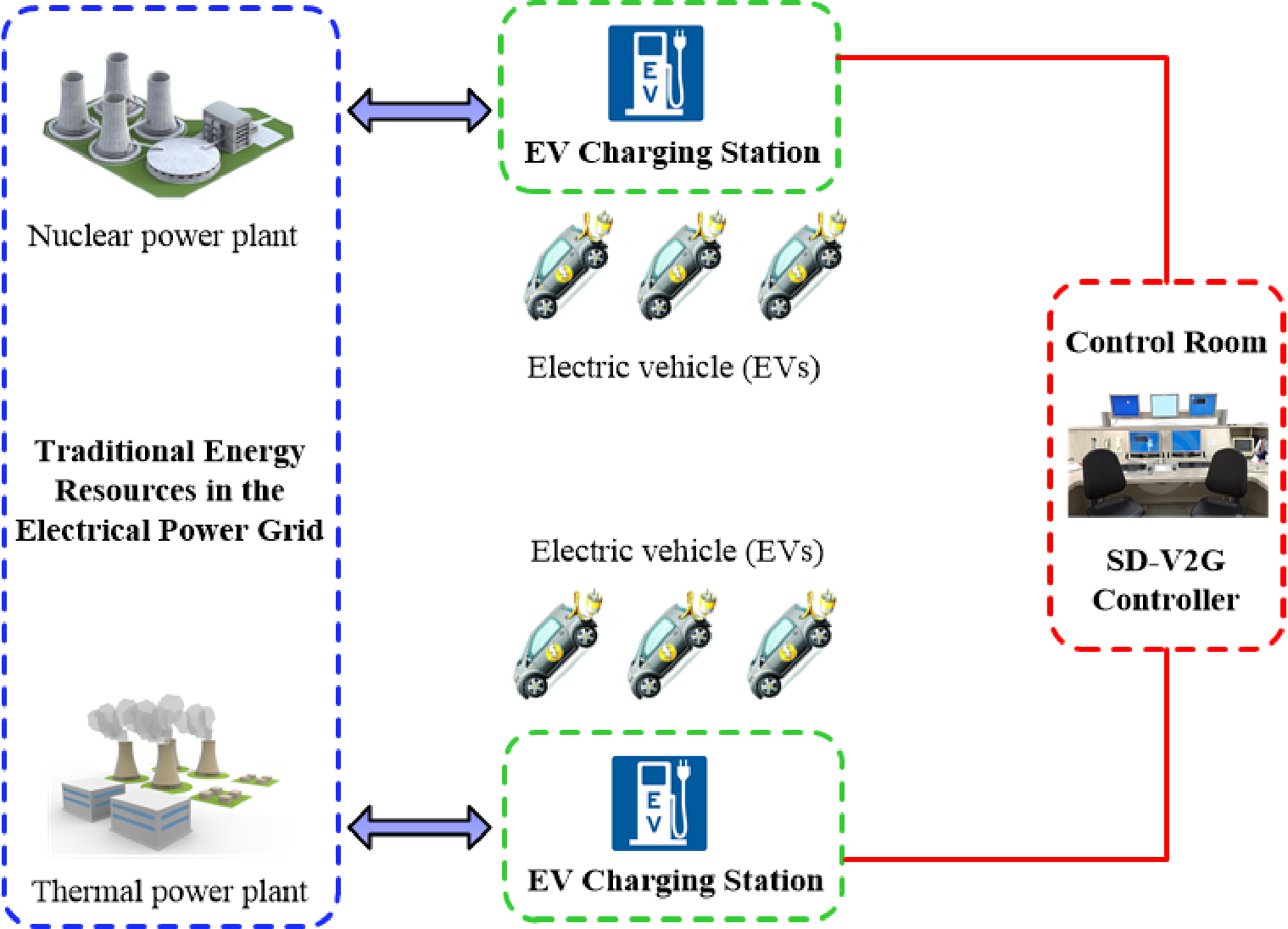}
\caption{\drmub{V2G} network in which electric vehicles (EVs) are connected with the charging stations to charge themselves or inject back the stored energy when required by the grid. The V2G charging stations are programmed through SD-V2G controllers\protect\cite{Li17sensors}.} 
\label{fig:sdv2g}
\end{figure*}

\subsection{\drmub{SDN-based SGC} architectures applied to Vehicles}
\label{sdnvehar}

In this sub-section, we discuss \drmub{SDN-based SGC} architectures applied to vehicles.

\subsubsection{\drmub{SDN-based SGC} applied to \drmub{V2G}}

In a Vechicle-to-Grid (V2G) setting, electric vehicles can also provide or inject energy into the \drmub{SG}. This injected energy by the EVs can be helpful to stablize the \drmub{SG} or to meet the peak energy requirements. \mub{In the context of V2G, Chekired~\cite{Chekired18tii} proposes a decentralized cloud SDN architecture for the SG.} In order to deal with peak loads, authors suggested to use a dynamic pricing model for EV charging and discharging. There were two proposed architectures for V2G: (a) Fog SDN, and (b) Cloud SDN. In the fog SDN architecture, each microgrid will be equipped with a decentralized cloud computing feature, while in the cloud SDN architecture, cloud data center will be used and a centralized SG controller will be present on the cloud. The authors proposed to use SDN in both the cloud and fog architecture but no specific details about the SDN implementation was discussed. Simulations were performed in Matlab and SUMO and it is shown that the proposed cloud SDN architecture further improves the SG stability. 

Another SDN-V2G architecture is proposed in~\cite{Zhang16sege}. The main purpose of this SDN-V2G architecture is to deal with security attacks. SDN-V2G architecture consists of application layer, control layer, and the data layer. The application layer supports different applications such as alarm processing, traffic control, maintenance, and secure communication. The control layer contains the SDN based V2G controller which is responsible for tasks such as data calculation, data management and link information collections. Finally, the data layer will be capturing the sensed data. The application layer will be connected with the control layer through \drmub{NBI} REST. The control layer will be connected with the data layer through \drmub{SDI} extensible messaging and presence
protocol (XMPP). We provide more details on this architecture from the security perspective later in Section~\ref{vsec}.

\mub{Li et al.~\cite{Li17sensors} proposes a battery status sensing for software defined \drmub{V2G} architecture (SDV2G).} There are two building blocks on SDV2G architecture. The first one is SD-V2G controller and data plane. The SD-V2G controller contains the \drmub{NBI} and facilitate different applications such as node configuration, flow table generation, and topology discovery. SD-V2G connects with the data plane through \drmub{SBI} (SNMP server). The data plane contains smart EVs and sensors. More details on this architecture can be found in Section~\ref{mcasv2}.

\mub{Similar to Li's work, Sun et al.~\cite{Sun17commag} proposes a software defined electric vehicle charging network architecture.} The proposed architecture contains three planes: the application plane, the control plane, and the physical plane. The application plane will be connected with the control plane through NBI, while the control plane will be connected with the physical plane through SBI. The application plane has two components namely management center and the intelligent decision making center. The management center contains applications such as generation management, data management, and communication resource management, while the intelligent decision making center will help to make \mub{intelligent decisions} through knowledge base using learning algorithm. The control plane will manage two main entities namely information control and energy control. And finally, the physical plane will have all the physical devices (EVs, IEDs, sensors) and grid infrastructure. It is demonstrated through a case study that the SDN based vehicle charging station provides more flexibility over the non-SDN based counterpart, and showed that grid operation cost and demand settling time can be reduced significantly. 

An SDN-based PEV integration framework for SG is proposed in~\cite{Chen17network}. The proposed framework is composed of two tiers. The upper tier focus on primary feeder level applications, while the lower tier focuses on secondary feeder level applications. The primary feeder level applications includes data monitoring and \drmub{DR} for large size commercial and industrial SG components. The secondary feeder level applications includes data monitoring and \drmub{DR} for small sized \drmub{SG} components. More precisely, the upper tier of this framework will provide a more general global view of the SG system, while the lower tier provides more microscopic view of the SG system. Authors then provided a case study to demonstrate the effectiveness of using SDN in PEV integrated with SG. By using SDN framework, the V2G operations can be fully supported by exploiting passing by PEVs.

\mub{A three plane architecture consisting of management plane, control plane, and the data plane for SDN-based green V2G considering energy management is proposed in~\cite{Hu18Commag}. The data plane is further logically divided into stationary data plane and dynamic data plane. The management plane consists of service provide and network manager. The control plane consists of SDN controller for both data and energy control. And the data plane consists of all the devices used for data transmission. With the help of proposed SDN-based green V2G architecture, it is shown that the average energy utilization is higher than non-SDN based architecture. }

\mub{A software-defined Wi-V2G architecture is proposed in~\cite{Wang18Itsm}. Wi-V2G is basically an IEEE 802.11 \drmub{Wi-Fi}-based multihop
wireless mesh network based on WiLD nodes which uses IEEE 802.11n wireless cards to guarantees long-distance communication. The performance of Wi-V2G is evaluated through Exata emulator and it is shown that the design of Wi-V2G ensures low infrastructure cost, capable to handle high mobility of EVs, and transmission of state of EV charging stations in real-time. On top of these advantages, Wi-V2G exploits the use of SDN technology to ensure flexible, robust, and centralized network management.}

\subsubsection{\drmub{SDN-based SGC} applied to \drmub{G2V}}
\label{sdnsgg2v}

\mub{Nafi et al.~\cite{Nafi16icspcs} proposed a \drmub{G2V} load management scheme applied to SG NAN in conjunction with SDWSN.} Authors proposed to use SDN to enable adaptive energy supply, however, the details regarding the SDN controller type and SDN switch type is not provided. Though the authors evaluated their proposed load management scheme in Castalia but fine grained details about SDN is not provided. By using the SDN-based \drmub{G2V} load management scheme, load management is achieved. The proposed scheme helps to optimize the daily load curve by using valley filling technique. In valley filling technique, the peak load is shifted to non-peak hours so that utility do not generate extra power at peak hours.

\subsection{\drmub{SDN-based SGC} architecture applied to Cellular Networks}
\label{sdnlte}

An \drmub{SDN-based SGC} architecture for cellular networks is presented in~\cite{Rubaye17wcom}. For \drmub{SG} modernization, wireless communication is essential and it is well established in the literature that the spectrum is a scarce resource and has been already over-crowded. Thus, the telecommunication industry is facing shortage in wireless radio spectrum. For utilities, it is not a good option to deploy their own telecommunication infrastructure, instead, the utilities are now moving towards using resources of existing telecommunication industry on lease basis. This will further make the wireless radio spectrum over-crowded. To address this challenge, the third generation partnership project (3GPP) has proposed to use licensed and unlicensed bands and they called it LTE's licensed assisted access (LAA). In LTE-LAA, two wireless radio spectrum bands will be used: the LTE licensed band and \drmub{Wi-Fi} 5-5.8GHz unlicensed band. Thus, with LTE-LAA, the \drmub{SG} utilities will be able to use both the licensed spectrum band of LTE and unlicensed band of \drmub{Wi-Fi}.

\mub{To manage resources for utilities and perform spectrum assignment for utility equipment, Rubaye et al.~\cite{Rubaye17wcom} proposed to use SDN technology.} More precisely, two algorithms were proposed. The first algorithm is for SDN controller, which performs spectrum management and the second algorithm is used for avoiding interference based upon predefined SDN policies. With these two algorithms, the SDN controller will help the utilities to monitor their communication network efficiently and to assign the spectrum while considering the interference and capacity demands. Furthermore, this SDN controller will keep history of the spectrum usage for making intelligent spectrum assignment decisions.

\begin{figure*}[t] \centering
\includegraphics[width=1\textwidth]{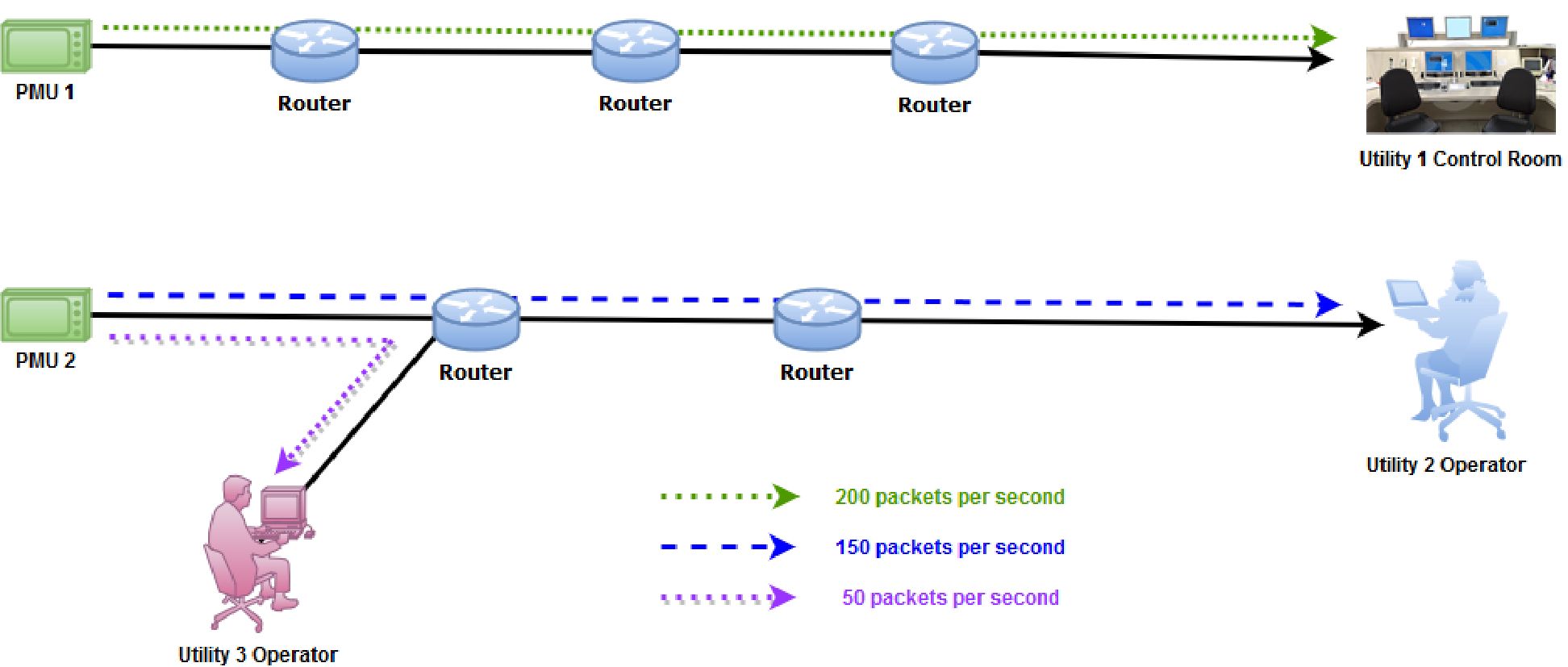}
\caption{Traffic with different data rate is generated in the PMU network\protect\cite{Goodney13smartgridcom}. Each utility requires PMU data with different packet rates, thus generating multirate traffic in PMU network.} 
\label{fig:pmumulti}
\end{figure*}

\subsection{\mhr{Summary and Lessons Learned}}

\mhr{In this section, we have surveyed architectures of SDN-based SGC. These architectures generally applied to optical networks, WSNs, vehicles, and to the cellular networks. If we compare the existing work on architectures applied to SDN-based SGC, we found that less number of architectures was proposed for cellular networks and WSNs as compared to vehicles and optical networks. We observed that different architectures were proposed for V2G and G2V, however, these architectures were all dealing with PEVs. Therefore, it will also be a good line of research to further investigate about proposing single architecture which unifies V2G and G2V paradigms together. We envisage that there is a need to propose new architectural frameworks for energy Internet, information centric networking, and virtual private networks.}

\section{\drmub{Routing Schemes for SDN-based SGC}}
\label{rout}

In this section, we discuss routing schemes for \drmub{SDN-based SGC}. Table~\ref{tab:routmulti} provides the summary of routing schemes proposed for SDN-based \drmub{SGC}. \drmub{We classify the existing routing schemes from the perspective of message delivery i.e., unicast routing, multicast routing, and broadcast routing. In unicast routing, message is delivered from a single source to a single destination. In multicast routing, message is delivered to a group of nodes. While in broadcast routing, message is delivered to all the nodes.} Below we discuss each of them in detail.

\subsection{\drmub{Unicast Routing Schemes for SDN-based SGC}}
\label{unrtr}

\subsubsection{\drmub{Unicast Routing Schemes for General SG Communication}}
\label{rgen}

Unicast Routing schemes for general SG communication has been considered in few works. \mub{Zhang et al.~\cite{Zhqng16icsn} calculates the shortest restore path for quick fault diagnosis.} Moreover, end-to-end path restoration is achieved through SDN-based controller (NOX controller). The proposed scheme was evaluated under normal scenario, after the expansion of network, and after the occurrence of the fault.  

\mub{Double constrained shortest path (DCSP) routing algorithm has been proposed by Zhao et al.~\cite{Zhao2016}. Authors then compared it with Dijkstra algorithm.} By considering test case for New England Test Power System using Mininet and Floodlight controller in a SDN-enabled SGC, authors demonstrated that DCSP achieves higher bandwidth and lesser delay compared to Dijkstra. In fact, SDN can help to find shortest path due to Global SDN controller. A similar study is done in~\cite{Alharbi16smartgridcom} in which a routing path searching algorithm is proposed. Authors \drmub{implemented} their proposed algorithm in Floodlight controller and compared it with shortest path routing algorithm.  

\mub{General SG applications produce huge amount of data being generated from various SG devices. Communicating simply this data without processing it will result in the wastage of resources and link congestion. Thus, in order to avoid these issues, efficient data transmission schemes have been proposed in the literature. One such a scheme to forward reduced data named as empirical probability-based routing algorithm (EPCS) for SDN controller is proposed in~\cite{Kaur18Tkde}. In the proposed EPCS routing scheme, first the generated big data is reduced by applying Tensor-based data reduction method to reduce its dimensionality. Then, this Tensor-based reduced data is intelligently routed through SDN controller by choosing those paths which has reduced data or maximum likelihood for scheduling. The demonstrated results are encouraging as EPCS increases the throughput and minimize the delay. One of the unique features of EPCS routing scheme is that it considers QoS requirements. Moreover, EPCS selects optimal routes while considering latency, load, bandwidth, and channel capacity.}

\mub{A three-stage routing strategy for \drmub{SG} in wireless mesh network setting is proposed in~\cite{Patil16nca}. In the proposed routing strategy, first connection is established between the controller and switch. Then path optimization is performed between the controller and the switches. And finally, routing among SDN enabled switches is performed for making efficient routing decisions. To evaluate the performance of the proposed routing protocol, latency has been used as a metric by using POX controller, Mininet, and NS-3 based simulations. This SDN based three stage routing strategy is superior to non-SDN counterpart approaches in a sense that it helps the SDN controller to select globally optimal routes with less delay while incurring less overhead and less manual intervention to configure routers.}

\mub{Information centric networking (ICN) can be applied to SDN-based \drmub{SGC} systems. One advantage of using ICN in SG systems is that it enables large scale exchange of data without knowing the IP address of the host. This may result in a secure SG by avoiding attacks which are invoked for a particular server or host having specific IP address. In ICN paradigm, caching routers are responsible to deliver data to the requesting SG entities, however, there may be a time synchronization problem between the caching routers and SG devices. In order to mitigate this time synchronization problem in caching routers, SDN-based on-path time synchronization (SD-OPTS) strategy is proposed in~\cite{Han17Globecom}. SD-OPTS shares the time stamps to all on-path caching routers to mitigate sync errors and decrease delay in caching routers. Simulation result in ndnSIM simulator, which is basically built on top of NS-3, demonstrates the effectiveness of SD-OPTS.}

\subsubsection{\drmub{Unicast Routing Schemes for Industrial Systems}}
Industrial control systems such as \drmub{SG} based on Internet of Things (IoT) devices are evolving day by day. \mub{In this context, Rubaye et al.~\cite{Rubaye17iot} focused on establishing routes for control data of SG through SDN controller (OpenVswitch is used in fact for evaluation purpose).} The proposed SDN framework can reset SDN switch or re-write the traffic to handle the faults. 

\mub{{\it OptimalFlow} which is basically serves as a novel control plane in SDN enabled industrial systems has been proposed by Genge et al.~\cite{Genge16ifip}.} With the help of integer linear programming (ILP), authors proposed to use shortest routing path and then harmonize different traffic flows. Single domain and multiple domain SDN scenarios were considered while evaluating the proposed control plane. \drmub{The proposed unicast routing algorithm select shortest path for the flow between source and destination pair having links with the largest capacities.}

\subsubsection{\drmub{Unicast Routing Schemes for Substation Communication}}
\mub{Germano et al.~\cite{Germano15im} proposes to use SDN in SCADA system in the substation environment.} The goal is to provide privacy against the eavesdropping attack in which an intruder may try to fully capture the communication data. Privacy is achieved by using multipath routing enabled by SDN which frequently shifts the communication routes of the SCADA system. In this manner, communication between the SCADA devices will be performed over more than one communication route and thus mitigating the affect of eavesdropping attack. More precisely, authors proposed an SDN based anti-eavesdropping algorithm for SCADA system in which multipath routes are selected. Evaluations were performed using POX OpenFlow controller and Mininet and shows that in the presence of the proposed multipath routing approach, eavesdropping was difficult to achieve.

\subsection{\drmub{Multicast Routing Schemes for SDN-based SGC}}
\label{multiSDN-based SG}

In multicast routing, information is communicated to several users or a group of users. In the context of \drmub{SG}, multicast routing has been used to disseminate time critical information such as control commands or measurement data from the PMUs. In wide area monitoring system, PMUs measure voltage and current information. This measured information is then multicast to the control center for immediate action. \mub{Similarly, utilities can multicast to a large number of consumers} to switch-off their appliances at peak hours to manage the power level in the grid~\cite{Li11tsg}. Inside the substation, multicast communication can also be done to disseminate emergency alerts across substation LANs. Multicasting can also be used for firmware updates required by a subset of smart meters~\cite{Tonyali17infocom}. In SDN-based \drmub{SGC}, multicast routing schemes have been applied to \drmub{V2G} networks, \drmub{PMU} networks, and substation communication.

\subsubsection{\drmub{Multicast Routing Schemes for V2G}}
\label{mcasv2}

Electric vehicles (EVs) will be the essential part of the future \drmub{SG}. EVs will not only reduce the CO$_2$ emissions in the environment but they can also inject back the stored energy to stabilize the power grid when required~\cite{Mouftah18wiley}. EVs can be considered as moving power plants which may help to supply power to areas where energy is required. In order to make the \drmub{V2G} network functional, several sensors are required to install at charging stations (CS) as well on the EVs itself. These sensors help the EVs owners to monitor the charging conditions of their vehicles. Moreover, by sharing these sensing information to the V2G network, the overall power grid can be stabilized. For instance, at peak hours, the V2G network can schedule the charging of EVs so as to meet the demand supply of power grid. However, in order to do so, the V2G needs battery status of EVs and their state of charge (SOC). In this manner, the V2G can instruct the EVs through multicasting and implement the regulations pertaining SOC. \mub{Li et al.~\cite{Li17sensors} proposes a battery status sensing software-defined multicast (BSS-SDM) scheme.} In this scheme, the SDN-based centralized controller is responsible for the monitoring and control of sensors of EVs (as depicted in Fig.~\ref{fig:sdv2g}). The SDN-based controller is chosen to support V2G multicast as it will support mobility of EVs as well as dynamic configuration can be easily done. Authors used simple network management protocol (SNMP) as \drmub{SBI} protocol and demonstrated through extensive simulations that their proposed BSS-SDM scheme can reduce average delay time cost of V2G operations.

\subsubsection{\drmub{Multicast Routing Schemes for PMU Network}}
\label{mcapmu}

Wide area mointoring of electrical power grid is an essential task in \drmub{SG} which is normally done through \drmub{PMUs}. PMUs provide current, frequency, and voltage information of the power system in real time~\cite{Ghosh13tii}. The information regarding wide area situational awareness recorded by PMUs need to be transmitted to multiple clients (utilities) at high speed, i.e., 20-200 ms delay with 99.99\% reliability, for control and protection of the power system. In fact, a utility can subscribe to several PMUs of their own grid and of the other neighbouring grids to monitor its overall health. In this manner, multiple utilities can collect data from PMUs at the same time, leading to multicast traffic generation in the PMU network. Not necessarily that every utility requires the PMU's data at same interval, some utilities may require more frequent data than the others, thus generating multirate traffic, as shown in Fig.~\ref{fig:pmumulti}. Unsuccessful timely delivery of this PMU data may result in power system failure. \mub{Goodney et al.~\cite{Goodney13smartgridcom} exploits SDN's {\it programmability} feature to deal with multicast and multirate features of PMU traffic.} Authors showed that by using OpenFlow, PMU's traffic can be optimally transmitted without making any changes.

\mub{Gyllstrom et al.~\cite{Gyllstrom14sgc} proposes APPLESEED}, which is composed of PCOUNT, MULTICAST RECYCLING, PROACTIVE, REACTIVE, and MERGER algorithms, designed specifically to meet the traffic requirements of PMU data. To accurately detect the packet loss inside the network, PCOUNT is proposed. MULTICAST RECYCLING algorithm is used to reduce the control traffic by the switches. PROACTIVE and REACTIVE algorithms are responsible for creating multicast backup trees for OpenFlow switches. Finally, MERGER is an optimization of PROACTIVE and REACTIVE algorithms. 
All these algorithms will run on OpenFlow enabled SDN switches. Authors evaluated their algorithms using Mininet.

\subsubsection{\drmub{Multicast Routing Schemes for Substation Communication}}

\mub{Pfeiffenberger et al.~\cite{Pfeiff15ntms} focuses on the use of SDN-based multicasting in substation environment.} Since reliability is required in substation communication, therefore, authors concentrated on fault tolerance and efficient delivery of multicast traffic. In order to achieve less packet loss, authors adopted to use fast-failover groups features of OpenFlow. Fast-failover groups features in OpenFlow are designed to detect and overcome port failures~\cite{openff}. 

\mub{Lopes et al.~\cite{Lopes15ac} considers IEC 61850 substation SGC environment.} Authors \drmub{proposed} SMARTFlow, an architecture in which two algorithms are proposed. The first \mub{algorithm calculates the} multicast tree and routes, while the second algorithm performs fault detection and restore the communication network. The goal of these two algorithms are to disseminate the GOOSE and SV messages on priority and in case of network failure, reconfigure the flow entries. SMARTFlow is evaluated using POX OpenFlow protocol and tested in Mininet environment. It is showed through extensive simulations that \mub{SMARTFlow generates less than 40\% overall network load compared} to a typical switch.

\subsubsection{\drmub{Multicast Routing Schemes for AMI Networks}}
\label{raminet}

In \drmub{HANs}, each house is equipped with an smart meter enabling two-way communication between the consumers and the utilities. These smart meters monitor the energy consumption of the household devices and then communicate the energy consumption to the utilities \mub{through a centralized aggregator} in neighbourhood area networks (NANs)~\cite{ArendP12}. NANs aggregators are then connected through advanced communication infrastructure to the utilities. In this manner, an \drmub{AMI} network is constituted which is responsible for real time prizing and demand side management~\cite{Finster15comst}. 

Routing for NANs in AMI network is extensively reviewed in~\cite{Ramirez15jnca}, however, the routing protocols discussed there are not related to SDN. To the best of our knowledge, there is only two studies~\cite{Kim15smartgridcom,Montazerolghaem18IoT} which considered multicast routing for SDN-based AMI network, however~\cite{Kim15smartgridcom} supports both multicast and broadcast routing (which we discuss in Sec.~\ref{mbrsami}). 

\mub{Due to new SG applications, the data generated by smart meters need to be communicated at smaller intervals to the smart meter data management system (SMDMS) maintained by the utilities. Moreover, as the SG is rolling out, the number of smart meters deployed in a particular region will increase dramatically. As a consequence, huge amount of data need to be reliably communicated to/from the SMDMS to the consumers. This may lead to congestion and load balancing problem at the intermediate nodes in SMDMS. Thus, OpenAMI, a global load-balancing multicast routing algorithm is proposed in~\cite{Montazerolghaem18IoT}. OpenAMI calculates the optimal routing path through Lagrangian relaxation-based aggregated cost (LARAC) scheme which relies on Dijkstra algorithm. \drmub{Best paths were selected to forward AMI data to two SMDMS.} A testbed is deployed using OpenvSwitch, Floodlight, and Kamailio to evaluate the performance of OpenAMI under various conditions such as constant offered load, variable offered load, and in cloud computing environment where virtual SMDMS are present. It is demonstrated that OpenAMI outperforms other existing approaches in terms of end-to-end delay and delivery ratio using load using offered by SDN technology in the entire AMI network.}

\begin{table*}[t]\footnotesize
\centering
\caption{Comparative view of routing schemes from the perspective of message delivery for \drmub{SDN-based SGC}.}
\label{tab:routmulti}
\begin{tabular}{|p{0.8cm}|p{1.3cm}|p{2.5cm}|p{2cm}|p{4.2cm}|p{4.3cm}|}
\hline
\centering
\bfseries Scheme & \bfseries \mub{Publication Year} & \bfseries \drmub{Type of Routing from Message Delivery Perspective} & \bfseries Type of SG Communication & \bfseries Metrics Evaluated & \bfseries Simulation Tool  \\
\hline
\cite{Zhqng16icsn} & \mub{2016}     & \multirow{9}{*}{} & General SG Communication & CPU utilization, end-to-end delay & OpenvSwitch, OpenFlow NOX controller \\ \cline{4-6}
\cite{Zhao2016} & \mub{2016}     &  & General SG Communication & Bandwidth, delay & Mininet, Floodlight, New England Test Power System testbed \\ \cline{4-6}
\cite{Alharbi16smartgridcom} &  \mub{2016}    &  & General SG Communication & Throughput, utilization of links & Mininet, OpenvSwitch, Floodlight controller \\ \cline{4-6}
\mub{~\cite{Kaur18Tkde}} &  \mub{2018}    &  & \mub{General SG Communication} & \mub{Throughput, delay, bandwidth usage} & \mub{`R' Programming and Matlab} \\ \cline{4-6}
\mub{~\cite{Patil16nca}} &  \mub{2016}    & \drmub{Unicast} & \mub{General SG Communication} & \mub{Latency} & \mub{Mininet, NS-3, OpenFlow POX controller} \\ \cline{4-6}
\mub{~\cite{Han17Globecom}} &  \mub{2017}    &  & \mub{General SG Communication} & \mub{Delay} & \mub{OpenFlow, NS-3 based ndnSIM} \\ \cline{4-6}
\cite{Rubaye17iot} &  \mub{2017}    &  & Industrial System & End-to-end delay, latency, data flow traffic & OpenvSwitch, OpenDayLight \\ \cline{4-6}
\cite{Genge16ifip} & \mub{2016}    &  & Industrial System & Average link load, No. of migrated and disconnected flows, throughput & OptimalFlow OpenFlow controller, Mininet, Floodlight controller \\ \cline{4-6}
\cite{Germano15im} &  \mub{2015}   &  & Substation Communication & Amount of exposed communication, packet loss, control traffic & OpenFlow, Mininet \\ \cline{4-6}
\hline
\cite{Li17sensors} & \mub{2017}  & \multirow{8}{*}{}  & V2G & Average delay cost, sensing time during EV aggregation & Not specified \\ \cline{4-6}
\cite{Gyllstrom14sgc} & \mub{2014}  &  & PMU Network & Loss estimates, No. of control messages & Mininet, POX OpenFlow controller \\ \cline{4-6}
\cite{Goodney13smartgridcom} &  \mub{2013} &  & PMU Network & Software and hardware switching latency & NOX OpenFlow controller, DETER tested \\ \cline{4-6}
\mub{~\cite{Montazerolghaem18IoT}} &  \mub{2018}  &  & \mub{AMI Network} & \mub{Request delivery ratio, end-to-end delay, established sessions, CPU consumption} & \mub{OpenvSwitch, Floodlight, Kamailio} \\ \cline{4-6}
\cite{Zheng16smartgridcom} & \mub{2016}  & \drmub{Multicast} & Optical Network & Bit error rate, end-to-end delay & Physical layer and data link layer simulations were performed \\ \cline{4-6}
\cite{Dorsch14smartgridcom} &  \mub{2014}   &  & Substation Communication & Data rate, delay & OpenFlow, OpenvSwitch \\ \cline{4-6}
\cite{Pfeiff15ntms} & \mub{2015}   &  & Substation Communication & No. of links for faultless solution & OpenFlow, Mininet \\ \cline{4-6}
\cite{Lopes15ac} & \mub{2015}   &  & Substation Communication & Delay, control load, overall network load & OpenFlow, Mininet \\ \cline{4-6}
\mub{~\cite{Zhou18Tii}} & \mub{2018}     & & \mub{Substation Communication} & \mub{Packet delivery success rate, throughput, memory usage} & \mub{OpenFlow, NS-3} \\ \cline{4-6}
\hline
\cite{Alishahi2016p2p} &  \mub{2016}    & \multirow{2}{*}{}  & General SG Communication & End-to-end delay, packet loss & Riverbed \\ \cline{4-6}
\cite{Kim15smartgridcom} &  \mub{2015}   & \drmub{Multicast/Broadcast} & AMI Network & Packet delivery ratio, end-to-end delay, average hop distance & OpenFlow Floodlight, NS-3 \\ 
\hline
\end{tabular}
\end{table*}

\subsubsection{\drmub{Multicast Routing Schemes for Optical Networks}}
\label{rton}

For substation communication, an hybrid electro-optical Ethernet based communication network structure is proposed by~\cite{Zheng16smartgridcom}. More precisely, authors used SDN in optical Ethernet network for routing and SDN center switch is used to reduce address learning process and routing time.

\subsubsection{\drmub{Multicast Routing Schemes for Substation Communication}}
\label{rutsbs}
SDN based routing schemes for substation communication is proposed in~\cite{Dorsch14smartgridcom},~\cite{Pfeiff15ntms}, ~\cite{Germano15im} and~\mub{~\cite{Zhou18Tii}}. In~\cite{Dorsch14smartgridcom}, different SG scenarios of transmission and distribution power grid were considered. For instance, fast recovery of \drmub{SG} communication and ensuring SG QoS were considered. Authors considered \drmub{SDN-based SGC} and compared it with traditional routing and QoS. Alternate routing paths, route reservation, and re-routing is done through SDN-based controller and significant gains were achieved in terms of SG traffic reliability. \mub{Pfeiffenberger et al.~\cite{Pfeiff15ntms} uses the fast-failover groups features of OpenFlow protocol to provide reliable routes.} Authors also demonstrated that their solution can be extended to a more general case where more than one link failures occur. \drmub{Minimum Spanning Tree broadcast routing has been used to approximate Steiner tree for connecting group of nodes having a minimum weight.}

\mub{For communication between the substations and the control centre, multiple communication paths may available to deal with link failures. However, to get advantage of using any of paths available, a multi-path routing algorithm is necessary. OpenFlow-based non-disjoint Path Aggregation Mechanism (OPAM) multi-path routing algorithm is one such algorithm proposed in~\cite{Zhou18Tii}. OPAM has two components: one is running at the SDN controller and the other one is running at the switches. The component running on controller is based on Multi-Path Subgraph Generation Algorithm (MPSGA) and the algorithm running on switches is Packet Aggregation and Forwarding Module (PAFM). Through extensive simulations performed in NS-3, it is demonstrated that OPAM outperformed existing multi-path routing algorithms in terms of packet delivery success rate, throughput, and memory usage of buffered packets.}

\subsection{\drmub{Routing Schemes Supporting Both Multicast and Broadcast for SDN-based SGC}}
\label{rtmb}

\subsubsection{\drmub{Multicast and Broadcast Routing Schemes for General SG Communication}}
Routing Protocol for Low Power and Lossy Network (RPL) has been used as a de-facto standard routing protocol in SGC~\cite{Khan17commag}. A virtual version of this RPL protocol i.e., optimized multi-class RPL (OMC-RPL) based on SDN and virtualization, has been proposed in~\cite{Alishahi2016p2p}. \drmub{OMC-RPL supports both multicast routing and broadcast routing as it is based on RPL.} It is demonstrated through extensive simulations that less message exchange for different traffic classes in SG is achieved through the concept of \mub{virtualization. }

\subsubsection{\drmub{Multicast and Broadcast Routing Schemes for AMI Networks}}
\label{mbrsami}
In ~\cite{Kim15smartgridcom}, SDN has been used to optimize for resource-constrained AMI devices. Authors proposed CoAP-SDN which is in fact a routing strategy. \drmub{CoAP-SDN used both multicast and broadcast schemes to transmit topology information announcement (TIA) messages in the network.} Authors evaluated their approach using IEC 61850 standard and showed through NS-3 based simulations that CoAP based SDN outperformed traditional SDN in terms of control message overhead, reliability, and QoS for large scale SDN-based AMI network. 

\subsection{\mub{Summary and Lessons Learned}}
In this section, \mub{we have provided a detail description} of routing schemes proposed for SDN-based \drmub{SGC}. The literature review reveals that multicast routing is only applied to V2G, PMU, and substation communication networks. However, there are several other applications in \drmub{SG} where multicasting can be used. For instance, firmware updates can be communicated to the group of smart meters in AMI network through reliable multicasting~\cite{Tonyali17icccn}. Another application area is \drmub{DR} management with multicasting in cellular networks such as LTE~\cite{Saxena15icc}. Thus, the thorough examination of other application areas in SDN-based \drmub{SGC} is an important direction for future research. Traditionally, multicasting is done through IP multicast, however, with the incorporation of software defined capability in switches, there is a need to further investigate about protocols and techniques which enable security aware SDN enabled IP multicast for \drmub{SG} network applications~\cite{Islam18comst}.

One of the unique feature of applying SDN to SGC is that it helps the communication network management team to introduce additional functionalities in the communication devices (e.g., switches), thus, enabling the power system to deal with spontaneous failures and converting the SG system into a self-healing SG. This openness to new functionalities is very helpful in routing as well, as in traditional routing, routes were to be configured manually. However, \drmub{SDN-based SGC} offers flexibility to create redundant routes, re-routing, and route reservation dynamically. Additionally, through SDN, there will be ease in the deployment of routing protocols for NANs in particular~\cite{Ramirez15jnca} and SG in general. Moreover, the SDN controller can have the global picture of whole network and make the routing decisions in a more intelligent manner by considering the condition of the whole communication network. However, this global SDN controller may consider as a single point of failure, thus, leading to communication break-down for routing in the \drmub{SDN-based SGC}. Therefore, further investigation is required to handle SDN controller failure problem. Moreover, cross-layer SDN enabled routing protocols is another venue for future research direction which need to be explored in the context of SG.

\section{Security and Privacy in \drmub{SDN-based SGC}}
\label{secur}
Security and privacy issues for SDN are discussed in~\cite{Rawat17comst,Dargahi17comst,Scott16comst,Gonzalez16splitech}, while security and privacy related work for SG are surveyed in~\cite{YaQS12,LiXL12,ArendP92}. In contrast to these works, for SDN-based \drmub{SGC}, lot of work has been done on security and privacy aspects~\cite{Martins16andescon,Raw16pesgm,Silva16gc}. \drmub{In the right most branch of Fig.~\ref{fadvsdnsg},} we show the classification of security and privacy schemes for SDN-based \drmub{SGC}. We classify these security and privacy schemes according to substation communication, AMI network and when they are applied to V2G and WLANs.

Though there has been works which discusses security issues for SG such as a European project, Future Internet Smart Utility Services (FINESCE), which discuss in detail the security requirements for smart utility networks~\cite{Pozuelo16icsgs}. But, security issues arise due to SDN integration in SG has not been well discussed. 

\drmub{SGs} are critical infrastructure and they should be resilient under malicious attacks situations or when accidential failure occurs. A systematic review is conducted in~\cite{Dong15cpss} to show the effectiveness of using SDN when SG is under attack. Authors explored how SDN can increase the resilience of the SG by adopting SDN. Moreover, authors also discussed extra risks bring to the SG by the incorporation of SDN. It has been highlighted that \drmub{SDN-based SGC} security threats \mub{belongs to three major} classes: (a) SDN controller or applications running on SDN controller may get compromised, (b) SG power device may get compromised. Such a device can be a relay, a SCADA slave, or a remote terminal unit (RTU), and (c) SDN switch may get compromised. These three major attack classes can be applicable to any SG architecture. For instance, if a SDN switch of a V2G network may get compromised than it has a less severe impact than a SDN switch of a substation may get compromised. 

Several different scenarios including threat models and the required security techniques for \drmub{SDN-based SGC} have been discussed in~\cite{Akkaya15lcn}. In this paper, authors \drmub{looked} the security and privacy issues in \drmub{SDN-based SGC} from another perspective. Authors basically \drmub{classified} \drmub{SDN-based SGC} security threats into four major classes. The first one is ``Method Specific Threats'' in which the way of execution of threats is considered. The second one is ``Target Specific Threats'' in which any particular types of devices are targeted. The third one is ``Software Specific Threats'' in which the software itself will be the target of the malicious user. Finally, the fourth one is ``Identity Specific Threats'' in which the association of the attacker is considered, whether it is within the association or outside the association. Below we discuss the security and privacy techniques proposed for \drmub{SDN-based SGC} in more detail.

\subsection{Security Schemes Applied to Substation}
\label{subst}

\subsubsection{IDS}
The centralized SDN controller can be a single point of failure in the context of \drmub{SDN-based SGC} (cf. Section~\ref{sdnctrlfail} for more details). The control center in the substation will run the SDN controller and the SCADA master. \mub{Both of them} can be vulnerable to security threats. For instance, an attack can led the dis-functionality of SCADA master or SDN controller. Moreover, an attack can also harm to the application running on SCADA master or SDN controller. The result will be the disruption in control commands by the SCADA master or control traffic by the SDN controller. Similarly, OpenFlow switches may also get compromised. In this context, a framework for intrusion detection system (IDS) for substation in SG is proposed in~\cite{Ghosh17icdcsw}. In this study, authors suggested to use global SDN controller as well as local SDN controller besides security controllers to protect the SG. Furthermore, a \mub{light-weight} identity based cryptography approach is used for the protection of SG. The SDN based IDS framework will be used for monitoring the commands issued by the SCADA master as well as the local SDN controller so that the global state of the SG is monitored. Authors tested their proposed IDS framework on IEEE 37-bus in Mininet. RYU SDN controller and OpenFlow switches were used during the evaluation process. 

\mub{A network based \drmub{IDS} architecture (NIDS) for SDN-based SCADA systems has been proposed by Silva et al.~\cite{Silva16compsac}.} One-class classification (OCC) algorithm is basically proposed in NIDS. The NIDS architecture contains one main control center, eight distribution substations, four intermediate control center, and plenty of field devices. OpenFlow SDN controller is used for evaluation purpose. It is demonstrated that the OCC algorithm achieves 98\% accuracy to detect the intrusion in SCADA system.

A anti-eavesdropping scheme is also proposed for SCADA system in which secure communication is achieved by using multipath routing~\cite{Germano15im} (cf. Section~\ref{rutsbs} for more details).

\subsubsection{Link Flood Attack}
\mub{Link flood attacks in the context of substation communication network has been considered by Maziku et al.~\cite{Maziku17icnc}.} In this regard, authors \drmub{proposed} a security score model based on SDN. For the experimentation purpose, authors used GENI testbed (SDN enabled windmill collector substation) in which IEC 61850 standard messages were considered. In fact, SDN is used to handle the congested link. The OpenFlow controller helped to easily enforce the QoS policies and capable to identify heaviest flows and busiest communications links at real time.

\subsubsection{\mub{Authentication}}

\mub{In a \drmub{SDN-based SGC} environment, link failure may occur, resulting in the activation of link failure restoration mechanism by the SDN controller. In this perspective, an authentication scheme for wireless backup links restoration has been proposed in which the impact of authentication during recovery phase was studied~\cite{Aydeger18ccnc}. The proposed authentication scheme considered three types of links: wired and wireless (\drmub{Wi-Fi} and LTE). The backup restoration links performance has been evaluated through FloodLight SDN controller, Mininet emulator, and NS-3 simulator. Both the proactive and reactive link failure detection schemes were compared using end-to-end delay, switching delay, and packet loss performance metrics. This authentication scheme is particularly designed for general purpose MMS traffic standard and it is demonstrated that the proposed authentication scheme can still meets the end-to-end delays (100 ms for MMS traffic) of SG communication requirement. }

\subsection{Security Schemes Applied to AMI and PMU Networks}
\label{amipm}

\subsubsection{AMI Networks}
\mub{For the protection of data of AMI network, Irfan et al.~\cite{Irfan15cit} proposes a security architecture.} The proposed security architecture is based on SDN. To be more precise, authors used Flowvisor SDN controller which served as the virtualization entity for slicing purpose and helps to ensure different security parameters such as authorization, authentication, and confidentiality. Furthermore, LTE is used for the sending of smart meter data which is then compared with AES-128 encrypted metering data sent by the SG controller. Mininet, and NS-3 is used for evaluation \mub{purpose.} In the considered scenario, eight smart meters are connected with five SDN switches.

\mub{Zhang el al.~\cite{Whang17mis} proposes an efficient and privacy-aware power injection (EPPI) security scheme for AMI networks.} One of the key feature of EPPI is that it easily detects the replay attacks (a type of attack in which the intruder captures the record of valid packets and replay it). Moreover, EPPI generates message authentication codes for preserving the privacy of the customers in AMI Network. However, the weak aspect of this article is that no fine grain details regarding the SDN architecture is provided. 

\mub{An SDN-based threat detection mechanism for AMI network is proposed in~\cite{Chi14Secureware}. In the proposed threat detection mechanism, OpenFlow protocol has been modified. More precisely, traditional \drmub{IDS} Snort, which is a signature-based threat detection mechanism, is integrated with OpenFlow switches. However, one may argue that how this SDN-based threat detection mechanism is superior to non-SDN-based approach (traditional Snort deployment)? The answer to this question is that in traditional Snort deployment, the Snort system checks all the in-coming traffic according to the pre-defined rules and directs the firewall to block any suspicious traffic. While in the proposed SDN-based Snort deployment, the OpenFlow controller will forward all in-coming traffic for the analysis of Snort, which will then inform the OpenFlow controller to drop the suspicious traffic. There are few advantages of this approach: first, the administrator has fine-grained control over traffic flows; second, dynamic redirection of forwarding paths can be easily done to secure AMI network; and finally, the threat detection is distributed, resulting in an efficient detection of the threat over the AMI network.}

\mub{To secure smart meter data, the concept of local controller is introduced in~\cite{Aujla18Tii}. These local controllers will be deployed in a particular region where they collect metering data from smart meters using wireless links. Multiple local controllers will then communicate this metering data to control centre through fibre or power line communication. All these metering data and control units will be managed by a SDN controller through OpenFlow and virtual OF-switches. Cuckoo filter-based fast forwarding scheme is used by the control plane to forward the metering data. Then this data is secured through attribute based encryption scheme. On top of this, Kerberos is used to authenticate peer devices. Parameters such as throughput and load were tested in a SG testbed which outperformed existing approaches.}

\subsubsection{PMU Networks}
\label{pmunet}
Security schemes for SDN based PMU networks are proposed in~\cite{Lin18tsg,Jin17tsg}. \mub{Jin et al.~\cite{Jin17tsg} (which is basically an extension of the work by the same authors~\cite{Lin18tsg}, see Section~\ref{sfheal} for more details) offers the integration of SDN technology with microgrid.} A microgrid is in fact an independent power grid which facilitates the local community to meet their energy needs. Microgrid may or may not be coupled with the main utility grid. In this study, authors focused on security and reliability that SDN brings to the microgrid. Authors basically developed a SDN testbed for microgrid evaluation which is DSSnet. In fact, DSSnet was developed in~\cite{Hannon16cpads} and then extended by authors further in this paper. This testbed is deployed at Illinois Institute of Technology (IIT), USA. This DSSnet testbed consists of OpenDSS power system simulator, ONOS SDN controller, and Mininet emulator to evaluate microgrid operations. Authors then \drmub{evaluated} SDN based self-healing approach over this DSSnet testbed and compared their approach with spanning tree protocol (STP) and rapid spanning tree protocol (RSTP). Authors also evaluated specification based intrusion detection in PMU network.

\subsection{Security Schemes Applied to Different Network}
\label{diffnet}

\subsubsection{V2G}
\label{vsec}

\mub{A software defined Vehicle-to-Grid (SDN-V2G) architecture has been proposed by Zhang et al.~\cite{Zhang16sege}.} In this architecture, authors \drmub{tried} to handle different types of attacks which may faced by a SDN-V2G architecture. These attacks can be on the utility server itself or it can be on the communication network of the utility or it can be on the SDN controller(s) or it can be on the charging stations or it can be on the devices or vehicles. Thus, considering all these attacks, authors \drmub{proposed} a security mechanism to deal with all SDN-V2G vulnerabilities. IEC 61850 standard is considered and the evaluations were performed using Mininet and Floodlight. It is demonstrated through simulation results that SDN-V2G outperformed the existing V2G paradigm in terms of traffic load handling.

\subsubsection{VLANs}

\mub{Kim et al.~\cite{Kim14smartgridcom} proposes to use SDN based virtual utility network (SVUNs) architectural solution for M2M applications in SG.} Authors \drmub{suggested} to use SDN technology instead of IEEE 802.1Q for creating virtual LANs (VLANs), as IEEE 802.1Q cannot support high number of devices and only supports one time authentication for the M2M devices. This may result in security threats to the SG in case the M2M device get compromised. Therefore, SDN based virtual utility networks are good solution as the fine grained granularity provided by SDN to prevent security attacks even after the first time authentication of M2M devices. Authors evaluated their proposed SVUNs by creating a testbed having two Intel SDN switches using OpenFlow and showed that the increased end-to-end delay with increased number of M2M devices can be avoided.

\subsubsection{\mub{General SG Systems}}

\mub{\drmub{SDN-based SGC} communication systems are also connected with Internet to provide access to utilities. However, it may also endanger the \drmub{SDN-based SGC} systems in general and prone to different cyber attacks. These cyber attacks are Denial of Service attack, ARP poisoning attack, and host location identification attack. The affect of these cyber attacks on \drmub{SDN-based SGC} system has been studied in detail in~\cite{Ibdah17icecta}. It has shown that \drmub{SDN-based SGC} systems are vulnerable to attacks like these. These cyber attacks were tested on different SDN controllers such as RYU, Floodlight, and POX through Mininet based simulations. More precisely, it is concluded that POX controller did not perform well as compared to Floodlight controller and RYU controller in the presence of congestion attack.}

\mub{In a microgrid environment, an SDN-based architecture using power line communication (PLC) technique is proposed to handle cyber attacks~\cite{Danzi18arxiv}. In the proposed architecture, wireless channel is used for data exchange while PLC in a power grid is used to carry control information to share the state of the microgrid. The proposed SDN-based architecture has been evaluated using Matlab/Simulink simulator and shown that it is resilient to cyber attacks.}

\subsection{\mub{Summary and Lessons Learned}}

In this section, we have surveyed security and privacy schemes for SDN-based \drmub{SGC}. A handsome amount of work has been done on security in general. The existing work generally focus on security and privacy issues pertaining to substation communication, WLANs, V2G, and AMI and PMU networks. However, there are few security and privacy aspects which are missing and need attention from the research community. For instance, a malicious user may deploy a ``darknet'' inside the SG communication network using SDN's monitoring channels. Such a darknet may become too much vulnerable for SG communication network as it may send suspicious commands to SG devices. Moreover, there is very little work done on distributed SDN control compared to the SDN's single central controller~\cite{Bannour18comst}. A critical future security research area for \drmub{SDN-based SGC} is in the context of secure service provisioning~\cite{He12commag}. Services, for e.g., billing, account management, installation, and customer management along with their security requirements need to be considered while designing security protocols for \drmub{SDN-based SGC}.

\begin{table*}[!htbp]\footnotesize
\centering
\caption{Testbeds and Simulation Tools used for evaluating the performance of \drmub{SDN-based SGC}.}
\label{tab:prototy}
\begin{tabular}{|p{2.5cm}|p{9cm}|p{1.5cm}|p{2.5cm}|}
\hline
\bfseries Category & \bfseries Testbed/Simulation Tool Name & \bfseries  Reference & \bfseries \mub{Publication Year} \\
\hline
\multirow{11}{*}{}  & \multirow{5}{*}{} 	& \cite{Rubaye17iot} & \mub{2017} \\ \cline{3-4}
& 	&  \cite{Dorsch14smartgridcom}  & \mub{2014} \\ \cline{3-4}
& North Bound Interface	& \cite{Dorsch16smartgridcom}  & \mub{2016} \\ \cline{3-4}
& 	&  \cite{Genge16ifip} &  \mub{2016} \\ \cline{3-4}
& 	& \cite{Sun17commag} &  \mub{2017} \\ \cline{2-4}
APIs & \multirow{6}{*}{}  & \cite{Wang16infocom} &  \mub{2016} \\ \cline{3-4}
&   & \cite{Rubaye17iot}  &  \mub{2017} \\ \cline{3-4}
&   & \cite{Dorsch14smartgridcom}  &  \mub{2014} \\ \cline{3-4}
& South Bound Interface  & \cite{Genge16ifip} &  \mub{2016} \\ \cline{3-4}
&   & \cite{Sun17commag}  &  \mub{2017} \\ \cline{3-4}
&   & ~\cite{Li17sensors} &  \mub{2017} \\ \cline{3-4}
\hline
\multirow{8}{*}{} & GSN testbed/Prototype	&  \cite{Ngyuen13ic} &  \mub{2013} \\ \cline{2-4} 
 & \multirow{2}{*}{}GENI~\cite{Berman15comnet}	& \cite{Sydney14comnet}  &  \mub{2014} \\ \cline{3-4} 
 & 	& \cite{Maziku17icnc}  &  \mub{2017} \\ \cline{2-4} 
 & Kansas State University as a SG prototype	&  \cite{Sydney14comnet} &  \mub{2014} \\ \cline{2-4} 
Testbeds & CloudLab, and RTPIS Lab	&  \cite{Narayanan16psc} &   \mub{2016} \\ \cline{2-4}
 & New England Test Power System	&  \cite{Zhao2016} &  \mub{2016} \\ \cline{2-4}
 & 24 nodes US and 28 node EU physical topology	& \cite{Rastegarfar16icnc} &  \mub{2016} \\ \cline{2-4}
 & FPGA based SDN control module	& \cite{Zheng14icocn} &  \mub{2014} \\ \cline{2-4}
\hline
Emulators & SDN-based Network Emulator	&  \cite{Hannon16cpads} &  \mub{2016} \\
\hline
\multirow{50}{*}{} & \multirow{2}{*}{}RYU~\cite{ryu} & \cite{Ren17ae} &  \mub{2017} \\ \cline{3-4}
&  & \cite{Lopes17im} &  \mub{2017} \\ \cline{2-4}
 & \multirow{3}{*}{} &  \cite{Rubaye17iot}  &  \mub{2017} \\ \cline{3-4}
 & OpenDayLight~\cite{opendaylight} &   \cite{Aydeger15nfvsdn}  &  \mub{2015} \\ \cline{3-4}
 &  &  \cite{Gonzalez16splitech} &  \mub{2016} \\ \cline{2-4} 
 & \multirow{10}{*}{}	&  \cite{Irfan15cit}  &  \mub{2015} \\ \cline{3-4}
 &  &  \cite{Lopes17im} &  \mub{2017} \\ \cline{3-4}
 &  &  \cite{Zhang16sege}  &  \mub{2016} \\ \cline{3-4}
 &  &  \cite{Guo16wimob}  &  \mub{2016} \\ \cline{3-4}
 & Mininet~\cite{mininet}  &  \cite{Aydeger15nfvsdn} &  \mub{2015} \\ \cline{3-4}
 &  &  \cite{Aydeger16icc}  &  \mub{2016} \\ \cline{3-4}
 &  &  \cite{Zhao2016}  &  \mub{2016} \\ \cline{3-4}
 &  &  \cite{Gyllstrom14sgc}  &  \mub{2014} \\ \cline{3-4}
 &  &  \cite{Molina15caee}  &  \mub{2015} \\ \cline{3-4}
 &  &  \cite{Germano15im} &  \mub{2015} \\ \cline{2-4}
 & \multirow{11}{*}{}	&  \cite{Dorsch14smartgridcom} &  \mub{2014} \\ \cline{3-4}
 & 	&  \cite{Sydney14comnet} &  \mub{2014} \\ \cline{3-4}
 & 	&  \cite{Li16sege} &  \mub{2016} \\ \cline{3-4}
 & 	&  \cite{Sydney13tsg} &  \mub{2013} \\ \cline{3-4}
 & 	&  \cite{Aydeger16icc} &  \mub{2016} \\ \cline{3-4}
 & OpenFlow	&  \cite{Pfeiff15ntms} &  \mub{2015} \\ \cline{3-4}
 & 	&  \cite{Ren17tsg} &  \mub{2017} \\ \cline{3-4}
 & 	&  \cite{Gyllstrom14sgc} &  \mub{2014} \\ \cline{3-4}
Simulation Tools, & 	&  \cite{Molina15caee} &  \mub{2015} \\ \cline{3-4}
SDN enabled & 	&  \cite{Hou17access}  &  \mub{2017} \\ \cline{3-4}
Controllers \& & 	&  \cite{Ngyuen13ic} &  \mub{2013} \\ \cline{2-4}
Switches & \multirow{2}{*}{}POX~\cite{pox} 	& \cite{Gyllstrom14sgc}  & \mub{2014} \\   \cline{3-4}
 & 	& \cite{Germano15im}  &  \mub{2015} \\ \cline{2-4}
 & Flowvisor	& \cite{Irfan15cit} &  \mub{2015} \\ \cline{2-4}
  & SUMO	& \cite{Chekired18tii}  &  \mub{2018} \\ \cline{2-4}
  & \multirow{2}{*}{}OPAL-RT	&  \cite{Ren17ae} &  \mub{2017} \\ \cline{3-4}
 & 	&  \cite{Ren17tsg} &  \mub{2017} \\ \cline{2-4}
 & THYME	&  \cite{Sydney13tsg} &  \mub{2013} \\ \cline{2-4} 
 & IEEE 30-bus and 118-bus system	&  \cite{Lin18tsg} &  \mub{2018} \\ \cline{2-4}
 & IEEE 37-bus system & ~\cite{Ghosh16cypss} &  \mub{2016} \\ \cline{2-4}
 & OptimalFlow	&  \cite{Genge16ifip} &  \mub{2016} \\ \cline{2-4}
 & \multirow{2}{*}{}MATLAB~\cite{matlab}	&  \cite{Chekired18tii}  &  \mub{2018} \\ \cline{3-4}
 & 	&   \cite{Pan16ccc} &  \mub{2016} \\ \cline{2-4}
 & IEEE 14-bus test network	&  \cite{Meloni17energies} &  \mub{2017} \\ \cline{2-4}
 & \multirow{2}{*}{}Openvswitch	&  \cite{Dorsch14smartgridcom} &  \mub{2014} \\ \cline{3-4}
 & 	&  \cite{Zhqng16icsn} &  \mub{2016} \\ \cline{2-4}
 & \multirow{3}{*}{}	&  \cite{Zhang16sege} &  \mub{2016} \\ \cline{3-4}
 & Floodlight~\cite{floodlight}	&  \cite{Dorsch16smartgridcom} &  \mub{2016} \\ \cline{3-4}
 & 	&  \cite{Zhao2016} &  \mub{2016} \\ \cline{2-4}
 & GRER 	& \cite{Hou17access} &  \mub{2017} \\ \cline{2-4}
 & \multirow{7}{*}{}	& \cite{Irfan15cit}  &  \mub{2015} \\ \cline{3-4} 
 & 	& \cite{Guo16wimob}   &  \mub{2016} \\ \cline{3-4} 
 & 	& \cite{Aydeger15nfvsdn}  &  \mub{2015} \\ \cline{3-4} 
 & NS-3~\cite{ns3}	& \cite{Sydney13tsg}   &  \mub{2013} \\ \cline{3-4} 
 & 	& \cite{Aydeger16icc}   &  \mub{2016} \\ \cline{3-4} 
 & 	& \cite{Pan16ccc}   &  \mub{2016} \\ \cline{3-4} 
 & 	& \cite{Kim15smartgridcom}  &  \mub{2015} \\ \cline{2-4} 
\hline
\end{tabular}
\end{table*}

\section{Open Issues, Challenges, and Future Research Directions}
\label{sec:openissues}

\subsection{\mub{Open Issues, Challenges, and Future Research Directions Related to SG Resilience}}

\subsubsection{\mub{Distributed SG Control Plane for SG Resilience}}
\mub{In Section~\ref{sdnctrlfail}, we have discussed that SDN controller is a single point of failure and it may severely suffer the SG communication resulting in loss of control to the network. This further aggravates the situation and may result in power outages~\cite{Mahmoodi16tett}. Generally speaking, a single SDN controller has been used in SG communication; however, one solution to mitigate this single point of failure or SG resilience issue is to use one or more SDN controllers (a.k.a distributed control plane) in SG environment. Distributed control plane is achieved by using cluster of SDN controllers to handle fast-failover. In this cluster, each SDN controller (replica) will manage few switches and also keeps information managed by other controllers. In case of a SDN controller's failure, the replica SDN controller will take over and resumes the operation of the network. However, keeping this information up-to-date will result in extra overhead on control channel as well as synchronization problem of different states of SDN controllers. Though some effort has been made to study the response time of controllers, and election of leader in the cluster in case of SDN controller failure~\cite{Sakic18Tnsm} but still there are few other future research directions such as reducing further the response time and control over that need to be explored.}

\drmub{Decentralization of SDN's control plane may also lead to consensus problem among the SDN controllers. In fact, in a distributed control plane, SDN controllers share their internal controller states to synchronize with each other. This can be considered as a fallback solution and helps to deal with another controller's failure problem. However, one problem arise in this situation is the lack of synchronization among the states of different SDN controllers (i.e., how to reach consensus among the SDN controllers). To deal with this consensus problem, consensus algorithms have been proposed and play an important role. One such a consensus algorithm is RAFT which is implemented by ONOS and OpenDaylight~\cite{Ongaro14usenix} and has been tested in industrial setting~\cite{Sakic18Tnsm}. Since cyberattacks are more common to critical infrastructure in industrial settings, therefore, consensus protocols need to be secure and trustable~\cite{Rehmani19Comst}. For instance, more recently a blockchain-based consensus protocol, Dueling DQL, for industrial IoT has been proposed~\cite{Qiu18iot}. Dueling DQL ensures trusted network-wide synchronization among SDN controllers using permissioned blockchains. However, measuring trust features among SDN controllers, also remains an open issue.}

\subsubsection{\mub{Link Failure Identification and Protection for SG Resilience}}
\mub{The protection and correct functioning of microgrid depends upon reliable, fast, and robust communication system~\cite{Habib18Tia}. Without such a communication system, there may be severe damage to microgrid hardware, thus leading to increase productivity cost and financial reparations. As discussed in Section~\ref{sgres}, SDN is a good choice to manage microgrid communication system; however, there are few research areas which are still undiscovered. For instance, there is a need to address the communication challenges of SDN in both islanded and grid connected modes of microgrid. From the SDN perspective in mircogrid, it is also required to investigate cases where SDN controller is able to reliably and quickly communicate the control information to relays to maximize safety to the equipment and also cope with system instability.}

\subsection{\mub{Open Issues, Challenges, and Future Research Directions Related to SG Stability}}

The future \drmub{SG} will facilitate the incorporation of distributed RERs at massive scale. With these small capacity but largely distributed RERs in the \drmub{SG}, the stable operation of the SG requires lot of coordination and control mechanism. Otherwise, some part of the SG may have excess energy, while the others may require lot of energy. In order to deal with this issue, the concept of virtual power plant (VPP) has been introduced in the literature. A \drmub{VPP} will logically aggregate distributed RERs and make them visible as a single entity to the rest of the power system. Definitely a VPP is based upon a centralized software which gathers these distributed RERs, therefore, assembling and dismantling these RERs quickly and create different VPP based upon geographical needs will be much easier with SDN~\cite{Maharjan15gc}. A VPP use case and opportunities of SDN usage has been described in detail by authors in~\cite{Zhang13cicsp}. Therefore, there is a need to further explore this direction of research and propose new algorithms and middlewares for VPP software~\cite{Zhang13cicsp}.

\subsection{\mub{Open Issues, Challenges, and Future Research Directions Related to SG Traffic Optimization}}

\subsubsection{In-band SDN Control for SG Communication and Traffic Optimization}
In SDN, the data plane is decoupled with the control plane. An interface is required to communicate between control plane and the data plane. The most famous and prominent protocol for such an interface is OpenFlow. OpenFlow is based on out-of-band network. An out-of-band network means that a separate dedicated communication link is required for control traffic between the controller and the switches. There are certain advantages of out-of-band communication for control traffic. For instance, in case of failure in data traffic paths, access to switches is even possible. Moreover, out-of-band communication is more secure as separate and dedicated communication links are used for controller/switch communication. However, there are few disadvantages of out-of-band network as well, such as dedicated communication link may be expensive and may be infeasible for large network topologies. Therefore in-band control traffic functionality is suggested in~\cite{Sharma16network}. In in-band control functionality, the control and data traffic will be using the same communication links, however, priority will be given to the control traffic. Using the control and data traffic on the same communication link (in-band network control), the control traffic may suffer and do not get the required preference and thus resulting in control traffic disconnection. This will seriously hampered the controller operations, such as load balancing, the establishment of new services, or commands passed by the controller to the switches. To address these challenges, authors in~\cite{Sharma16network} also proposed to include queuing functionality in OpenFlow protocol. Authors also proposed to use restoration and path protection in OpenFlow to provide resilient communication. However, there is still lot need to be done for in-band SDN control for SG communication networks.

\subsubsection{Network Coding for \drmub{SDN-based SGC} for Traffic Optimization}
Network coding is a technique in which packets are combined together at intermediate nodes to earn some gains (for e.g., reduction in packet transmission, security, and increase in throughput) and it has been applied to different communication networks such as cognitive radio networks~\cite{Naeem16comst,Farooqi14jnca}. Network coding has been applied to \drmub{SG} as well for various purposes such as to reduce the collection time of the data and to achieve reliability~\cite{Su2012bookchp,Phulpin11sgc,Prior14tsg}. 

\mub{Prior et al.~\cite{Prior14tsg} proposes to use network coding in the context of SG.} Authors basically considered an AMI network in which smart meters need to communicate periodically the information to the BS. The smart meter scenario contains lot of nodes and the type of generated packets are smaller in size. Considering this traffic requirement, authors used tunable sparse network coding concept in which at the beginning of communication, the network coding intensity is sparser and then with the passage of time, the coding density increase. By considering IEEE 37 and 123 node scenario, authors \mub{validated} their approach in NS-2. It is further demonstrated that with the proposed network coding approach, 100\% reliability is achieved and data gathering time is 10\% more faster.

With the ever increasing popularity of network coding, it has been suggested to become the part of 5G. \mub{Towards this goal, Hansen et al.~\cite{Hansen15commag} basically proposes to use network coding as a service for 5G using SDN.} The same concept can be further extended and one interesting research is to incorporate network coding as a service (NCaaS) for future SDN based \drmub{SG} communication. However, issues such as coding coefficients and variants of network coding and its impact on traffic conditions in SG need to be investigated in detail.

\subsection{\mub{Other Future Research Directions}}

\subsubsection{Securing SG Networks through SDN}
In Section~\ref{secur}, we have discussed that how SDN can be used to secure the \drmub{SG}. However, SDN also brings few security issues to the \drmub{SG} as well~\cite{Ali15tr}. In fact, securing \drmub{SDN-based SGC} itself is an area which require further investigation. For instance, despite the fact that SDN's control plane leveraged many advantages to the SG but this control plane may also get affected with attacks and become more vulnerable. Moreover, using a single centralized SDN controller in SG may become a single point of failure~\cite{don2015con}. One approach to handle this single point of failure of SDN controller is to decentralize the SDN control. Such a decentralization of SDN's control plane may alleviate the risk of Denial of Service (DoS) attack to a greater extent but another problem arises which is related with the intercommunication between SDN controllers. Imagine a situation in which a malicious node may act like a SDN controller in a decentralized SDN control plane to compromise the SG. Therefore, such types of new challenges in the context of SG need to be dealt with more attention~\cite{Bannour18comst}.

Another potential security threat is related with SDN based virtual network slices (see Section~\ref{motiv} for details) and SDN based virtual utility network~\cite{Kim14smartgridcom}. In both cases i.e., SDN based virtual network slices and virtual utility network, a SDN network hypervisor will play an active role and multiple virtual network slices or virtual SDNs (vSDNs) can be created through these hypervisors. Securing these SDN hypervisor is essential and further investigation is required to secure them in the context of \drmub{SDN-based SGC}.

\drmub{In a SGC system, several types of control protocols have been used such as IEEE 1815 (a.k.a., Distributed Network Protocol - DNP3), DNPSec, IEC 61850, and IEC 60870-5-101. However, these protocols are vulnerable to different types of attacks as highlighted in~\cite{Volkova18comst}. Designing security mechanisms to further strengthen these control protocols against various attacks is still an open area of research.}

\subsubsection{Routing Based on Content and Information Centric Networking in \drmub{SDN-based SGC}}
With the help of SDN, new services and applications can be easily incorporated to the \drmub{SG}~\cite{Youssef17iwqos}. One such as application of SDN in \drmub{SG} is to exploit information centric networking. For instance, the feasibility of ICN in the context of SG has been demonstrated on real power distribution network in Netherlands~\cite{Katsaros14network}. SDN can provide help in in-network content caching, routing based on content, and query/response through content-centric networking~\cite{Zhang13cicsp}. However, less work is done on the applications of content centric networking in SG from the perspective of SDN.

\subsubsection{Routing in Energy Internet and \drmub{SDN-based SGC}}
Energy Internet (EI) is a novel concept in which the goal is to connect the SG devices together for the effective management of power flows among different SG entities~\cite{Tenti17tsg,Bui12network}. Compared to the traditional Internet, there are routers and routing paths in the \drmub{EI} and the goal is to find optimal paths and routes for energy distribution~\cite{Wang17Tsc}. In this manner, efficiency is achieved in energy transmission. Similar to the traditional Internet where different local area networks (LANs) are connected to make a global Internet, we have Energy Local Area Networks (e-LANs) in the \drmub{EI} paradigm~\cite{Wang17tii}. In e-LANs, the energy routers are responsible for the management of energy flows among different SG devices and in this regard, different energy routing algorithms were proposed in the literature~\cite{Wang17tii}. Though there are some interesting primilinary restuls on the integration of \drmub{EI} and SDN~\cite{Zhang17eceiec,Hou17access,Chelmis16wfiot} but still there is lot of efforts required for the reliable and efficient working of SDN-based EI. \mub{Additionally, integrating real-time QoS capability to end-to-end traffic flows in EI and \drmub{SDN-based SGC} is also an interesting area to explore~\cite{Guck17tnsc}}.

\subsubsection{Interoperability and Standardization in \drmub{SDN-based SGC}}
In SG, different types of devices will interact with each other. Moreover, these devices belongs to different vendors, thus, SDN can help a lot in interoperability issues. For instance, in Neighborhood area \mub{network, devices} are based on proprietary wireless mesh network. NANs deployed in North America generally operates in sub GHz band which ranges from 902--928 MHz unlicensed band. NAN devices need to be interoperable in order to achieve economic benefits. For the interoperability of NAN devices, two NAN communication standards, IEEE 802.15.4g and IEEE 802.15.4e were published. It is envisage that an alliance like ZigBee Alliance or \drmub{Wi-Fi} Alliance is also required for NAN interoperability~\cite{Chang13wc}. In this context, SDN can help to make these NAN devices interoperable. \mub{Ustun et al.~\cite{Ustun16pecon} provides an interoperability framework for IEC 61850 based microgrid protection system in SG.} 

From the standardization perspective, lot of work has been done in SG communication~\cite{gun2011sma,ArendP11,ArendP20} and SDN alone. For instance, data-driven interoperability and communication-driven interoperability for SG has been proposed by~\cite{Kim17commag} but efforts are required to deal with \drmub{SDN-based SGC} interoperability issues so that utilities can easily adopt SDN technology~\cite{Jaebeom15aps}.

\subsubsection{Simulations Related Issues in \drmub{SDN-based SGC}}

In Table~\ref{tab:prototy}, we provide the summary of application programming interfaces (APIs), testbeds, simulation tools, SDN enabled controllers and switches, used for evaluating the performance of SDN-based \drmub{SGC} systems. It is clear from this table that there is very less work done on East Bound Interface (EBI) and West Bound Interface (WBI) in the context of \drmub{SDN-based SGC}. It is also evident that Mininet, OpenFlow, and NS-3 was the choice of most of the researchers to validate their algorithms and protocols for \drmub{SDN-based SGC}.

From the testbed perspective, though there are lot of testbeds available for evaluation of SDN~\cite{Huang17comst} and SG~\cite{Cintuglu17comst,Ollis16isgt} alone but less efforts are done on integrating these components together. For instance, an effort to integrate Mininet, PowerWorld, and \drmub{IDS} to check \drmub{SDN-based SGC} resilience has been done in~\cite{Dong15cpss}. Another such an effort is made in~\cite{Song16tecs} by proposing a hardware-software federated simulation platform in which cloud computing is used. However, a testbed combining all the features of power system, SDN, and SG need to be developed so that future protocols can be easily tested over them.

\mub{Substation configuration language (SCL) for IEC 61850 has been used to configure SG devices and equipment in a sub-station. However, there are number of security issues which need to be addressed in the context of SCL such as integration of cryptography, key management, device authentication and different types of attacks~\cite{Raw17pesgm}.}

Moreover, communication issues for distributed optimization in SG is another future research direction area which need to be investigated. In fact, distributed infrastructure is present in the SG environment and there is a need to communicate between the control centers of different utilities. For this purpose, inter-control center communications protocol (ICCP or IEC 60780-6/TASE.2) has been used in industry. \mub{Guo et al.~\cite{Guo17tii} studied this aspect and used OPNET modeler for evaluation purpose.} However, communication issues for distributed optimization need to be tightly coupled in simulation tools.

\subsubsection{Virtualization in SG through SDN}
\label{virtu}
Virtualization through \drmub{SDNs} has been a hot topic and a general discussion on it can be found in~\cite{Zhang17network}. In a similar fashion, virtualization has also been applied in SG through SDN~\cite{Kim14smartgridcom,Xin11smartgridcom,Meloni16energycon,Niedermeier16jnsm,Hammouri12iesgm,Maharjan15gc}.

For instnace, the concept of virtualization in substation communication is presented in~\cite{Dorsch14icc,Kurtz18netsoft}. By using the virtualization concept in substation, dynamic configuration and management case be easily performed. \mub{Dorsch et al.~\cite{Dorsch14icc} basically considers two type of traffic i.e., substation traffic based on IEC 61850 standard and traffic related to virtualization.} Extensive simulations for double star infrastructure, single Infrastructure, and dedicated infrastructure substation scenarios, were conducted in OPNET Modeler and it is concluded that SDN should be incorporated to use virtualization at network level.

Virtualization for wide area measurement system has been studied in which virtualized PMUs are introduced~\cite{Meloni16energycon}. Another work on virtuzliation of PMUs is presented in~\cite{Hammouri12iesgm}. SDN based virtual utility networks have been presented in~\cite{Kim14smartgridcom}. SDN based framework for network and power virtualization has been discussed in~\cite{Maharjan15gc}.

\mub{Meloni et al.~\cite{Meloni16icc} focuses on virtualizing PMUs and incorporating context awareness in them.} Basically, an IoT based state estimation system using cloud computing is proposed. It is demonstrated through extensive simulations that required QoS level can be easily achieved by using the proposed IoT cloud based system. A detailed discussion on network virtuzliation for \drmub{SG} has been presented in~\cite{Lv14systems}. \mub{Niedermeier et al.~\cite{Niedermeier16jnsm} uses network function virtualization (NFV) for creating a virtual AMI network.} A potential future research area is to consider the scalability and heterogeneity issue while designing virtualization based protocols for \drmub{SDN-based SGC}.

\section{Conclusion}
\label{sec:conclusion}
\drmub{SG} is composed of diverse set of electrical, control, and electronic devices ranging from \drmub{PMUs} to smart meters and from supervisory control and data acquisition system to complex power generating and dispatching units. The ultimate goal of all these devices is to provide services to the end users (either customers or utility operators). The success of \drmub{SG} underlie on an efficient, reliable, flexible, and globally managed communication system, which help to assist SG in providing these services in a timely manner. \drmub{SDN} meets these criteria and can help to form the basis for SG control. Moreover, by applying SDN in SG systems, efficiency and resiliency can potentially be improved. In this survey article, we have comprehensively covered the advantages of \drmub{SDN-based SGC} and presented its taxonomy. We have then discussed case studies on the use of SDN in SG. Routing schemes for \drmub{SDN-based SGC} were then classified and discussed in detail. Security schemes along with their classification are then provided. Finally, we have identified and discussed challenges, issues, and future research directions related to \drmub{SDN-based SGC} before concluding the paper.

\bibliographystyle{IEEEtran}


\begin{IEEEbiography}{Mubashir Husain Rehmani} (M'14-SM'15) received the B.Eng. degree in computer systems engineering from Mehran University of Engineering and Technology, Jamshoro, Pakistan, in 2004, the M.S. degree from the University of Paris XI, Paris, France, in 2008, and the Ph.D. degree from the University Pierre and Marie Curie, Paris, in 2011. He is currently working as an Assistant Lecturer at the Department of Computer Science, Cork Institute of Technology, Ireland. Prior to this, he worked as a Post Doctoral Researcher from 2017-2018 at the Telecommunications Software and Systems Group (TSSG), Waterford Institute of Technology (WIT), Waterford, Ireland. He served for five years as an Assistant Professor at COMSATS Institute of Information Technology, Wah Cantt., Pakistan.  He is currently an Area Editor of the IEEE COMMUNICATIONS SURVEYS AND TUTORIALS. He served for three years (from 2015 to 2017) as an Associate Editor of the IEEE COMMUNICATIONS SURVEYS AND TUTORIALS. Currently, he serves as Associate Editor of  IEEE COMMUNICATIONS MAGAZINE, \textit{Journal of Network and Computer Applications (Elsevier)}, and the \textit{Journal of Communications and Networks (JCN)}. He is also serving as a Guest Editor of \textit{Ad Hoc Networks journal (Elsevier)}, \textit{Future Generation Computer Systems journal (Elsevier)}, the IEEE TRANSACTIONS ON INDUSTRIAL INFORMATICS, and \textit{Pervasive and Mobile Computing journal (Elsevier)}. He has authored/ edited two books published by IGI Global, USA, one book published by CRC Press, USA, and one book with Wiley, U.K. He received ``Best Researcher of the Year 2015 of COMSATS Wah'' award in 2015. He received the certificate of appreciation, ``Exemplary Editor of the IEEE COMMUNICATIONS SURVEYS AND TUTORIALS for the year 2015'' from the IEEE COMMUNICATIONS SOCIETY. He received Best Paper Award from IEEE ComSoc Technical Committee on Communications Systems Integration and Modeling (CSIM), in IEEE ICC 2017. He consecutively received research productivity award in 2016-17 and also ranked \# 1 in all Engineering disciplines from Pakistan Council for Science and Technology (PCST), Government of Pakistan. He also received Best Paper Award in 2017 from Higher Education Commission (HEC), Government of Pakistan.
\end{IEEEbiography}

\begin{IEEEbiography}{Alan Davy} received the B.Sc. (with Hons.) degree in applied computing and the Ph.D. degree from the Waterford Institute of Technology, Waterford, Ireland, in 2002 and 2008, respectively. Since 2002, he has been with the Telecommunications Software and Systems Group, originally as a student and then, since 2008, as a Post-Doctoral Researcher. In 2010, he was with IIT Madras, India, as an Assistant Professor, lecturing in network management systems. He was a recipient of the Marie Curie International Mobility Fellowship in 2010, which brought him to work at the Universitat Politecnica de Catalunya for two years. He is currently a Senior Research Fellow and the Research Unit Manager of the Emerging Networks Laboratory, Telecommunications Software and Systems Group. He is the coordinator of the EU H2020 FETOpen Project CIRCLE: Coordinating European Research on Molecular Communications. His current research interests include Virtualised Telecom Networks, Fog and Cloud Computing, Molecular Communications and TeraHertz Communication.
\end{IEEEbiography}

\begin{IEEEbiography}{Brendan Jennings} (M'05) received the BEng. and PhD degrees from Dublin City University in 1993 and 2001 respectively. He is the Head of Graduate Studies at Waterford Institute of Technology and is active as a Senior Researcher within the TSSG Emerging Networks Laboratory. He is a Principal Investigator with the SFI CONNECT Research Centre. He has spent periods as a Visiting Researcher with the KTH Royal Institute of Technology, Sweden, and in Dell EMC Research Europe, Ireland. He regularly serves on the organization and technical program committees of network and service management related conferences and will serve as Executive Chair for the IEEE ICC 2020 conference in Dublin. His research interests include network management, cloud computing, and molecular communications.
\end{IEEEbiography}

\begin{IEEEbiography}{Chadi Assi} (S'02-M'03-SM'08) received the B.Eng. degree from Lebanese University, Beirut, Lebanon, in 1997, and the Ph.D. degree from the City University of New York (CUNY) in 2003. He is currently a Full Professor with the Concordia Institute for Information Systems Engineering, Concordia University. Before joining Concordia University, he was a Visiting Researcher with Nokia Research Center, Boston, where he worked on quality of service in passive optical access networks. His research interests are in the areas of networks and network design and optimization. He was a recipient of the Prestigious Mina Rees Dissertation Award from CUNY in 2002 for his research on wavelength-division multiplexing optical networks. He is on the Editorial Board of IEEE COMMUNICATIONS SURVEYS AND TUTORIALS, the IEEE TRANSACTIONS ON COMMUNICATIONS, and the IEEE TRANSACTIONS ON VEHICULAR TECHNOLOGIES. His current research interests are in the areas of network design and optimization, network modeling, and network reliability.
\end{IEEEbiography}

\end{document}